\documentclass[journal]{IEEEtran}
\usepackage[pdftex]{graphicx}
\usepackage{cite}
\usepackage{verbatim}		
\usepackage{amsmath}
\usepackage{amssymb}
\usepackage{amsthm}
\usepackage{mathtools}	
\usepackage{grffile}	
\usepackage[tight,footnotesize]{subfigure}
\usepackage{microtype} 
\usepackage[misc]{ifsym}
\usepackage{color}
\usepackage{url}
\usepackage[ruled,vlined,linesnumbered]{algorithm2e}
\usepackage{bm} %
\usepackage{scalerel}

\usepackage[inline]{enumitem}
\usepackage{adjustbox}
\usepackage{tikz}
\usetikzlibrary{calc,intersections,arrows.meta,bending,patterns,angles,matrix}
\usepackage{pgfplots}
\usepgfplotslibrary{fillbetween}

\theoremstyle{remark}
\newtheorem{remarkx}{Remark}
\newenvironment{remark}
{\pushQED{\qed}\remarkx}
{\popQED\endremarkx}
\newtheorem{examplex}{Example}
\newenvironment{example}
{\pushQED{\qed}\examplex}
{\popQED\endexamplex}

\theoremstyle{definition}
\newtheorem{defn}{Definition}
\newtheorem{assump}{Assumption}

\theoremstyle{plain}
\newtheorem{theorem}{Theorem}
\newtheorem{lemma}{Lemma}
\newtheorem{coroll}{Corollary}
\newtheorem{prop}{Proposition}

\newtheorem*{claim*}{Claim}
\newtheorem{conj}{Conjecture}

\newcommand{\defeq}{:=} %

\newcommand{\dist}{{\rm dist}}
\newcommand{\dt}{\frac{{\rm d}}{{\rm d}t}}
\newcommand{\expo}[1]{{\rm exp} \left(#1\right)}				
\newcommand{\matr}[1]{\begin{bmatrix} #1 \end{bmatrix}}
\newcommand{\vmatr}[1]{\begin{vmatrix} #1 \end{vmatrix}}
\newcommand{\transpose}[1]{#1^\top}
\newcommand{\inv}[1]{#1^{-1}}	%
\newcommand{\diag}[1]{{\rm diag}\{ #1\}}

\newcommand{\chiup}{\raisebox{2pt}{$\chi$}}
\newcommand{\vf}{\chiup}
\newcommand{\identity}{{\rm id}}
\newcommand{\set}[1]{\mathcal{#1}}

\newcommand{\norm}[1]{\left\lVert#1\right\rVert}
\newcommand{\manifold}{\set{M}}
\newcommand{\mbr}[1][{}]{\mathbb{R}^{#1}}	%
\newcommand{\gradient}{\,\mathrm{grad}\,}
\newcommand{\orthoterm}{\bot_{\phi}}
\newcommand{\interior}{\mathrm{int\,}}
\newcommand{\scalemath}[2]{\scalebox{#1}{\mbox{\ensuremath{\displaystyle #2}}}}
\newcommand{\Rot}[2]{\mathrm{Rot}_{#1}(#2)}	%

\begin{document}
	\title{Topological Analysis of Vector-Field Guided Path Following on Manifolds}
	
	\author{Weijia Yao, \IEEEmembership{Member,~IEEE}, Bohuan Lin, Brian D. O. Anderson, \IEEEmembership{Life Fellow,~IEEE}, Ming Cao, \IEEEmembership{Fellow,~IEEE}%
		\thanks{W. Yao and M. Cao are with ENTEG, University of Groningen, the Netherlands.  B. Lin is with BI, University of Groningen, the Netherlands.  B. D. O. Anderson is with the Research School of Engineering, Australian National University, Canberra, Australia. \tt\small weijia.yao.new@outlook.com, m.cao@rug.nl, b.lin@rug.nl, brian.anderson@anu.edu.au. }
	}

	\maketitle
	
	\begin{abstract}
		A path-following control algorithm enables a system's trajectories under its guidance to converge to and evolve along a given geometric desired path. There exist various such algorithms, but many of them can only guarantee local convergence to the desired path in its neighborhood. In contrast, the control algorithms using a well-designed guiding vector field can ensure almost global convergence of trajectories to the desired path; here, ``almost'' means that in some cases, a measure-zero set of trajectories converge to the singular set where the vector field becomes zero (with all other trajectories converging to the desired path). In this paper, we first generalize the guiding vector field from the Euclidean space to a general smooth Riemannian manifold. This generalization can deal with path-following in some abstract configuration space (such as robot arm joint space). Then we show several theoretical results from a topological viewpoint. Specifically, we are motivated by the observation that singular points of the guiding vector field exist in many examples where the desired path is homeomorphic to the unit circle, but it is unknown whether the existence of singular points always holds in general (i.e., is inherent in the topology of the desired path). In the $n$-dimensional Euclidean space, we provide an affirmative answer, and conclude that it is not possible to guarantee global convergence to desired paths that are homeomorphic to the unit circle. Furthermore, we show that there always exist \emph{non-path-converging trajectories} (i.e., trajectories that do not converge to the desired path) starting from the boundary of a ball containing the desired path in an $n$-dimensional Euclidean space where $n \ge 3$. Examples are provided to illustrate the theoretical results.
	\end{abstract}
	
	
	\IEEEpeerreviewmaketitle
	
	\section{Introduction}
		
	In the path-following problem, a desired path is specified as a geometric object rather than a temporal function (c.f. trajectory tracking or output regulation problems \cite{aguiar2005path,aguiar2008performance}) such that the output of a system is required to converge to and propagate along the desired path eventually. To follow a desired path is fundamental in many applications, and there are different algorithms \cite{Sujit2014}, such as pure pursuit, line-of-sight (LOS), nonlinear guidance law (NLGL) and vector-field guided algorithms. These algorithms are often designed for general mobile robot \emph{kinematics models} (e.g., single-integrator \cite{Goncalves2010}, double-integrator  \cite{lawrence2008lyapunov}, unicycle \cite{yao2020auto,yao2021thesis}) and thus provide \emph{guidance} signals as input to the model, and the \emph{vehicle-specific} inner-loop \emph{dynamics control} is assumed to be sufficiently fast and accurate to track these guidance signals \cite{phillips2004mechanics,kapitanyuk2017guiding}. Thus one can consider the guidance layer (e.g., designing a guiding vector field) and the control layer (e.g., designing motor control algorithms) separately. Compared to trajectory tracking or output regulation algorithms, there is a separate interest in the study of path-following algorithms since they are more suitable for some applications, such as fixed-wing aircraft guidance and control \cite{rezende2018robust,lawrence2008lyapunov,Sujit2014}.
	
	Among various path-following algorithms, the vector-field guided path-following algorithms have been investigated extensively \cite{lakomy2017,Goncalves2010,nelson2007vector, Zhu2013, kapitanyuk2017guiding2}. In these algorithms, the desired path is usually implicitly or explicitly assumed to be a sufficiently smooth \emph{one-dimensional connected submanifold} in $\mbr[n]$ for regularity reasons. Thus, desired paths can be conveniently classified into two categories: those homeomorphic to the unit circle $\mathbb{S}^1$ if they are compact and those homeomorphic to the real line $\mbr[]$ otherwise \cite[Theorem 5.27]{lee2010topologicalmanifolds}. These algorithms utilize a sufficiently smooth vector field of which the integral curves are proved to converge to the desired path at least asymptotically. Moreover, once any point of a trajectory is on the desired path, the integral curves will keep evolving on the desired path \cite{Goncalves2010,lawrence2008lyapunov}. 
	
	Note that most, if not all, of the studies assume that the Euclidean space $\mbr[n]$ is the configuration space of the considered ordinary differential equation where the right-hand side is the designed vector field. Although the explicit expressions of vector fields $\vf$ vary in different studies (c.f. \cite{liang2016combined, lakomy2017,kapitanyuk2017guiding,lawrence2008lyapunov,Goncalves2010,Zhu2013,michalek2018vfo}), they generally consist of two components: a \emph{converging component} and a \emph{propagation component}. The converging component enables the integral curves of the vector field to approach the desired path, while the propagation component, orthogonal to the converging component, provides a tangential direction to the desired path, and thus helps the integral curves propagate along the desired path. The two-component structure of the vector field is intuitive and effective in solving the path-following problem. In \cite{Goncalves2010}, a time-varying component is added to the vector field to deal with a time-varying desired path.

	There are several advantages of the vector-field guided path-following algorithms. One of them is the removal of the condition requiring the initial point to be sufficiently close to the desired path, as otherwise commonly required by algorithms such as LOS \cite{fossen2003line}, among others \cite{Sujit2014}. In addition, the vector-field guided algorithms are shown to achieve the highest path-following accuracy with the least control efforts among several algorithms in benchmark tests to follow a circle and a straight line \cite{Sujit2014}. However, their major drawback is the existence of \emph{singular points} where the vector field becomes zero, and the consequences are: i) the integral curves of the vector field might only be extended in a finite time interval \cite{kapitanyuk2017guiding}; ii) normalization of the vector field, which is common in many practical applications \cite{yao2020auto,kapitanyuk2017guiding}, at the singular point is not well defined; iii) if there are singular points, then the algorithm does not have the desirable property of \emph{global convergence} to the desired path from any initial conditions, and the analysis becomes more difficult. Therefore, some existing studies either impose conservative assumptions ruling out attractive singular points or avoid providing detailed analysis for singular points \cite{Goncalves2010,rezende2018robust,liang2016combined, lakomy2017}. The study in \cite{kapitanyuk2017guiding} does not use these assumptions and shows that the integral curves of a 2D guiding vector field either converge to the desired path or the singular set, which consists of all singular points of the vector field. This dichotomy convergence result has been extended to a 3D vector field \cite{yao2018cdc,yao2020auto}. However, it is unknown whether this dichotomy convergence property holds for guiding vector fields defined on general manifolds, including $\mathbb{R}^m$ for $m>3$. %
	
	In the literature and in practice, vector-field guided path-following examples are typically illustrated with a desired path homeomorphic to the unit circle, such as a circle, an ellipse or a Cassini oval \cite{kapitanyuk2017guiding,lawrence2008lyapunov}. In these examples, singular points of the guiding vector field exist, which implies that global convergence to the desired path cannot be guaranteed. Therefore, a natural question arises:  \emph{do singular points of the guiding vector field always exist when the desired path is homeomorphic to the unit circle, and thus global convergence to the desired path is not guaranteed?} This question is simple when the configuration space is the two-dimensional Euclidean space (i.e., $\manifold=\mbr[2]$). Since the desired path is a closed orbit by construction, the Poincar\'e-Bendixson theorem concludes that there is at least one singular point of the vector field\footnote{Here, a singular point of the vector field happens to be an equilibrium point of the ordinary differential equation where the right-hand side is the vector field.} in the region enclosed by the desired path. Therefore, once a trajectory starts from the singular point, it stays there and thus global convergence to the desired path is not possible. However, the Poincar\'e-Bendixson theorem is only applicable for the planar case $\mbr[2]$, while the conclusion for the higher-dimensional case $\mbr[n]$,  where $n>2$, and general manifolds is still untreated. %
	
	\subsection{Contributions}
 	This paper extends the vector-field guided path-following algorithms to a general smooth Riemannian manifold $\manifold$ (see Sections \ref{subsec_gvf} and \ref{sec_calvf_manifold}). One reason to consider smooth manifolds $\manifold$ rather than the Euclidean space is the immediate relevance to potential applications, especially when one deals with mechanical systems \cite{bullo2019geometric}. For example, the control of revolute joint angles of a two-joint robot arm corresponds to the case where the manifold is a torus (i.e., $\manifold=\mathbb{T}^2=\mathbb{S}^1 \times \mathbb{S}^1$) in the joint space, and the attitude control of aircraft correspond to the special orthogonal group $SO(3)$ (see the fourth and fifth examples in Section \ref{sec_simulation}). We show that global convergence from any point in the torus $\mathbb{T}^2$ to the desired path $\set{P} \subseteq \mathbb{T}^2$ homeomorphic to the unit circle $\mathbb{S}^1$ is not possible, but this issue can be solved by lifting the torus $\mathbb{T}^2$ to its covering space $\mbr[2]$  (see Section \ref{sec_simulation}). 
	
	The first contribution of this study arises from the analysis related to the dichotomy convergence, stability and attractiveness for the vector field defined on general manifolds (see Section \ref{sec_dichotomy}). Specifically, we show that the dichotomy convergence property still holds for the vector-field guided path-following algorithms defined on the general smooth manifold $\manifold$. This means that trajectories either converge to the desired path on $\manifold$ or the singular set. This result is not only an extension of \cite{kapitanyuk2017guiding} which only considers $\mbr[2]$, but also plays an important role in the subsequent theoretical development (e.g., Corollary \ref{coroll_attra_P}, Corollary \ref{cor_nonattra_C}, Theorem \ref{thm_rn_global}). We also prove, under some mild conditions, the asymptotic stability of the desired path $\set{P}$ (i.e., Corollary \ref{coroll_attra_P}) and the non-attractiveness\footnote{A formal definition is provided subsequently, but note that it is possible for some trajectories starting outside a non-attractive set to approach that set in the limit, a fact which is perhaps counter-intuitive. Consider in $\mathbb{R}^2$ the system $\dot x_1=x_1, \dot x_2= -x_2$, with the origin as the singular set. } of the singular set $\set{C}$ (i.e., Corollary \ref{cor_nonattra_C}), which are highly desirable properties in any path-following algorithms. However, we show by an example that these two properties alone cannot guarantee the almost-global convergence property of the desired path (i.e., Example \ref{ex5}). All such analysis motivates the subsequent topological analysis.
		
	The second contribution is to answer with respect to initial conditions the question proposed above regarding the existence of singular points and the possibility of global convergence (see Section \ref{sec_singular_pts}). We first revisit a topological result (i.e., Lemma \ref{lemma_deform_retract}) revealing the relationship between a compact asymptotically stable embedded submanifold and its domain of attraction, and provide some interpretations along with an outline of our independent proof of this result (i.e., Remark \ref{remark_alternative_proof}). This reveals two essential elements behind the result: the regularity of the desired path (a compact asymptotically stable embedded submanifold) and the continuity of the \emph{first hitting time}. Notably, we show that when the configuration space is the $n$-dimensional Euclidean space (i.e., $\manifold=\mbr[n]$) and the desired path is homeomorphic to the unit circle, singular points of the vector field always exist, and it is impossible to enable trajectories to converge globally to the desired path from all initial conditions in $\mbr[n]$ (i.e., Theorem \ref{thm_rn_global}). 
	
	This impossibility result further motivates us to show the existence of \emph{non-path-converging trajectories} (i.e., trajectories that do not converge to the desired path), which is the third contribution. It turns out that every ball containing the desired path has at least one non-path-converging trajectory starting from its boundary in $\mbr[n]$ for $n \ge 3$ (i.e., Theorem \ref{thm_locating_diverging}). This topological result is related to the impossibility of global convergence to the destination point of integral curves of a feedback motion planner \cite[Chapter 8]{lavalle2006planning} in an obstacle-populated environment \cite{koditschek1990robot} (see Conjectures \ref{conjecture1} and \ref{conjecture2}). 
	
	\subsection{Paper structure}
	Section \ref{sec_vf_manifold} introduces the guiding vector field on a smooth $n$-dimensional Riemannian manifold for path following, and a concrete computation procedure of the vector field defined on manifolds is presented in Section \ref{sec_calvf_manifold}. Section \ref{sec_dichotomy} elaborates on the preliminary analysis of the convergence issues. %
	The main results are given in Section \ref{sec_singular_pts} regarding the existence of singular points, global convergence to the desired path and the existence of non-path-converging trajectories. Several examples are provided in Section \ref{sec_simulation} to verify the theoretical results. %
	
	\subsection{Preliminaries on topological and differential manifolds} \label{preliminary}
	Some basic concepts about topological and differential manifolds \cite{lee2015introduction, lee2010topologicalmanifolds} are explained here. Suppose $\set{X}, \set{Y}$ are topological spaces. A \emph{homeomorphism} (\emph{diffeomorphism} resp.) $f: \set{X} \to \set{Y}$ is a continuous (smooth resp.) bijection that has a continuous (smooth resp.) inverse. If there exists a homeomorphism between $\set{X}$ and $\set{Y}$, then $\set{X}$ and $\set{Y}$ are \emph{homeomorphic}, denoted by $\mathcal{X} \approx \mathcal{Y}$. 
	
	Let $f,g: \set{X} \to \set{Y}$ be continuous maps. \emph{A homotopy from $f$ to $g$} is a continuous map $H: \set{X} \times [0,1] \to \set{Y}$ such that $H(x,0)=f(x)$ and $H(x,1)=g(x)$ for all $x \in \set{X}$. If there exists such a homotopy, then $f$ and $g$ are \emph{homotopic}, denoted by $f \simeq g$. Let $h: \set{Y} \to \set{X}$ be another continuous map. If $f \circ h \simeq \identity_{\set{Y}}$ and $h \circ f \simeq \identity_{\set{X}}$, where $\identity_{(\cdot)}$ is the identity map, then $h$ is a \emph{homotopy inverse for $f$}, and $f$ is called a \emph{homotopy equivalence}. In this case, $\set{X}$ is \emph{homotopy equivalent to} $\set{Y}$.
	
	Let $\set{A} \subseteq \set{X}$, a continuous map $r: \set{X} \to \set{A}$ is a \emph{retraction} if the restriction of $r$ to $\set{A}$ is the identity map of $\set{A}$, or equivalently if $r \circ \iota_{\set{A}} = \identity_{\set{A}}$, where $\iota_{\set{A}}: \set{A} \to \set{X}$ is the inclusion map and $\identity_{\set{A}}$ is the identity map of $\set{A}$. In this case, $\set{A}$ is called a \emph{retract of $\set{X}$}. Furthermore, if $\iota_{\set{A}} \circ r$ is homotopic to the identity map of $\set{X}$ (i.e., $\iota_{\set{A}} \circ r  \simeq \identity_{\set{X}}$), then $r$ is a \emph{deformation retraction} and $\set{A}$ is called a \emph{deformation retract of $\set{X}$}. Equivalently, $\set{A}$ is a deformation retract of $\set{X}$ if there exists a homotopy $H: \set{X} \times [0,1] \to \set{X}$ that satisfies $H(x,0)=x$, $H(x,1) \in \set{A}$ for all $x \in X$ and $H(a,1)=a$ for all $a \in \set{A}$. In addition, if the homotopy $H$ is stationary on $\set{A}$, that is, the last equation is replaced by $H(a,t)=a$ for all $a \in \set{A}$ and all $t \in [0,1]$, then $r$ is a \emph{strong deformation retraction} and $\set{A}$ is called a \emph{strong deformation retract of $\set{X}$}. The space $\set{X}$ is called \emph{contractible} if the identity map of $\set{X}$ is homotopic to a constant map, or equivalently, if any point of $\set{X}$ is a deformation retract of $\set{X}$. This means that the whole space $\set{X}$ can be continuously shrunk to a point.
	
	Let $\set{M}$ and $\set{N}$ be smooth manifolds. The \emph{tangent space} $T_p \set{M}$ to $\set{M}$ at $p \in \set{M}$ is a linear space of maps $X_p: C^{\infty}(p) \to \mbr[]$ satisfying the linearity and product rules, where $C^{\infty}(p)$ denotes the set of smooth real-valued functions defined on an open neighborhood of $p$ \cite{nijmeijer1990nonlinear}. Other alternative and equivalent definitions of the tangent space are summarized in \cite[pp. 71-73]{lee2015introduction}, including the intuitive definition that the tangent space is, roughly speaking, a set of ``velocity vectors'' tangent to a curve on the manifold. Given a sufficiently smooth map $F: \set{M} \to \set{N}$, the \emph{tangent map (or differential) of $F$ at $p \in \set{M}$} is denoted by $F_{*p}: T_p \set{M} \to T_{F(p)} \set{N}$ and satisfies $F_{*p} (X_p) (f) = X_p (f \circ F)$ for $X_p \in T_p \set{M}$ and $f \in C^\infty(F(p))$. The notation $F_{*p}$ is used interchangeably with $d F_p$ or $dF\big|_{p}$. If the subscript $p$ is omitted, then it is a tangent map $F_{*}: T \set{M} \to T \set{N}$ defined at any $p \in \set{M}$, where the \emph{tangent bundle} $T \set{M}$ is the disjoint union of the tangent spaces at all points of $\set{M}$ (i.e., $T \set{M} \defeq \coprod_{p \in \set{M}} T_p \set{M}$), and $T \set{N}$ is defined analogously. If the tangent map $F_{*p}$ at $p$ is surjective, then $p$ is called a \emph{regular point} of $F$. If for every $p \in \inv{F}(q)$,  the tangent map $F_{*p}$ at $p$ is surjective, then $q \in \set{N}$ is called a \emph{regular value of $F$}. The definition of a tubular neighborhood is in \cite[pp. 137-139]{lee2015introduction}, and that of an embedded submanifold is in \cite[pp. 98-99]{lee2015introduction}.
	In this paper, unless otherwise specified, all the manifolds have no boundaries.

	\section{Guiding vector field for path following} \label{sec_vf_manifold}
	In the literature of path-following problems using a guiding vector field, usually the guiding vector field is defined on the Euclidean space $\mbr[n]$. In this section, we generalize the discussion to Riemannian manifolds and introduce notions (e.g., the distance) as generalized counterparts of those in the Euclidean space $\mbr[n]$ for the subsequent analysis. In other words, given a sufficiently smooth guiding vector field $\vf: \manifold \to T \manifold$, where $\manifold$ is a Riemannian manifold that also satisfies some regularity conditions presented later and $T \manifold$ is the tangent bundle \cite{lee2015introduction}, we investigate the solutions to the following autonomous ordinary differential equation:
		\begin{equation} \label{eq1}
			\dot{\xi}(t) = \vf(\xi(t)),
		\end{equation}
	where $\xi(t) \in \manifold$ usually corresponds to a physical quantity such as the position of a mobile robot, and $\vf(\xi(t))$ corresponds to the desired velocity of the robot. The manifold $\manifold$ is called the \emph{configuration space}. The guiding vector field defined on a Riemannian manifold is introduced in Section \ref{subsec_gvf} and some standard assumptions are presented in Section \ref{subsec_assump}.
	
	\subsection{Guiding vector fields on Riemannian manifolds}	\label{subsec_gvf}
	We introduce some concepts first. A Riemannian manifold is denoted by $(\manifold,g)$, where $g$ is the Riemannian metric \cite{lee2018introduction}. The distance between a point $p \in \manifold$ and a submanifold $\set{N} \subseteq \manifold$ is defined by $\dist(p, \set{N}) = \dist(\set{N}, p) \defeq \inf\{d(p, q) : q \in \set{N}\}$, where $d(\cdot, \cdot)$ is the Riemannian distance of two points in $\manifold$ \cite[p. 36]{lee2018introduction}. The distance between $\set{N}$ and another submanifold $\set{N}' \subseteq \manifold$ is defined by $\dist(\set{N}, \set{N}') = \dist(\set{N}', \set{N}) \defeq \inf\{d(r, q) : r \in \set{N}, q \in \set{N}'\}$. The tangent space of $\manifold$ at a point $p \in \manifold$ is denoted by $T_p \manifold$, and the length or norm of a tangent vector $v \in T_p \manifold$ is defined by $ \norm{v} = \langle v, v \rangle_g ^{1/2}$, where $\langle \cdot, \cdot \rangle_g$ is the inner product of tangent vectors in $T_p \manifold$. As a special case, if $\manifold=\mbr[n]$, then the Riemannian metric is replaced by the canonical Riemannian metric (i.e., Euclidean metric) on $\mbr[n]$, the Riemannian distance by the Euclidean distance and the inner product by the dot product.
		
	Suppose the configuration space of \eqref{eq1} is an $n$-dimensional smooth Riemannian manifold $(\manifold, g)$, which is oriented, connected and complete \cite[p. 340]{lee2015introduction}. Suppose a desired path $\set{P} \subseteq \manifold$ is described by the intersection of $(n-1)$ zero-level sets, that is,
	\begin{equation} \label{eqpath1}
	\mathcal{P} = \{ \xi \in \manifold : \phi_i(\xi)=0, \; i=1,\dots, n-1\},
	\end{equation}
	where $\phi_i: \manifold \to \mbr[]$, $i=1, \dots, n-1$, called \emph{surface functions} for convenience, are of differentiability class $C^2$. Such a geometric description of the desired path without explicit parametric form is common in the literature when $\manifold=\mbr[n]$ \cite{Goncalves2010,rezende2018robust,michalek2018vfo}. However, the set description of $\set{P}$ might not be desirable if no further restrictions are imposed; e.g., the set $\set{P}$ might be disconnected or even empty. Therefore, one usually needs to assume that $\set{P}$ is a connected one-dimensional submanifold in $\manifold$ such that it corresponds to a desired path in practice. One advantage of the level-set description \eqref{eqpath1} is that the distance of a point $\xi \in \manifold$ to the desired path $\set{P}$ can be \emph{approximated} by the value of $\norm{(\phi_1,\dots,\phi_{n-1})}$ under some mild assumptions to be proposed later. Thus one could avoid the explicit computation of the distance $\dist(\xi, \set{P})$, which is difficult even if the desired path is an ellipse in $\mbr[2]$. %
	
	For simplicity, we first briefly introduce the guiding vector field on $\mbr[n]$, and later extend it to the general manifold $\manifold$. The $n$-dimensional vector field $\vf: \mbr[n] \to \mbr[n]$ is \cite{yao2020auto}:
	\begin{equation} \label{gvfe}
		\vf(\xi) = \wedge(\nabla \phi_1(\xi), \dots, \nabla \phi_{n-1}(\xi)) - \sum_{i=1}^{n-1} k_i \phi_i(\xi) \nabla \phi_i(\xi), \tag{GVF-E} \\
	\end{equation}
	for $\xi \in \mbr[n]$, where $\nabla \phi_i: \mbr[n] \to \mbr[n]$ is the gradient of $\phi_i$, $k_i>0$ are constant gains, and $\wedge: \underbrace{\mbr[n] \times \dots \times \mbr[n]}_{n-1} \to \mbr[n]$ is the wedge product \cite[p. 355]{lee2015introduction}. Note that $\wedge(\nabla \phi_1(\xi), \dots, \nabla \phi_{n-1}(\xi))$ is orthogonal to each of the gradients $\nabla \phi_i(\xi)$ for $i=1,\dots,n-1$ \cite[Proposition 7.2.1]{galbis2012vector}.

	We explain the physical interpretation of the vector field in \eqref{gvfe}. As mentioned before, the vector field generally consists of two terms: the \emph{propagation term} and the \emph{convergence term}.  The propagation term $\wedge(\nabla \phi_1, \dots, \nabla \phi_{n-1})$ is orthogonal to each gradient vector $\nabla \phi_i$, and thus is tangent to each $c$-level surface described by $\{p \in \mbr[n] :\phi_i(p)=c\}$. This enables the trajectory to move along the intersection of these level surfaces, and especially move along the desired path when $c=0$. The forward or backward direction of the movement with regard to the desired path can be changed by switching the order of any two of the gradient vectors in the wedge product. The convergence term $- \sum_{i=1}^{n-1} k_i \phi_i \nabla \phi_i$ is a linear combination of the gradient vectors, with the state dependent ``weight'' $-k_i \phi_i$. Thus it provides a direction towards the intersection of the zero-level surfaces, which is the desired path $\set{P}$.
	
	Now we show how to generalize the former discussion from the Euclidean space $\mbr[n]$ to the Riemannian manifold $\manifold$. Specifically, the gradient $\nabla \phi_i(\xi)$ and the term $\wedge(\nabla \phi_1(\xi), \dots, \nabla \phi_{n-1}(\xi))$ for $\xi \in \mbr[n]$ will be replaced by their counterparts denoted by $\gradient \phi_i(\xi)$ and $\orthoterm(\xi)$ respectively for $\xi \in \manifold$. The Riemannian gradient $\gradient \phi_i(\xi) \in T_\xi \manifold$ is the tangent vector to $\manifold$ at $\xi \in \manifold$ such that for all tangent vectors $v \in T_\xi \manifold$, there holds
	\begin{equation} \label{eq_gradient}
		\langle \gradient \phi_i(\xi), v \rangle_g = d \phi_i \big|_{\xi} (v),
	\end{equation}
	where $d \phi_i \big|_{\xi}: T_\xi \manifold \to \mbr$ is the \textit{differential} of $\phi_i$ at $\xi \in \manifold$ \cite[p. 281]{lee2015introduction}. The other term $\orthoterm(\xi)\in T_\xi \manifold$ is the tangent vector such that for all tangent vectors $v \in T_\xi \manifold$, there holds
	\begin{equation} \label{eq_ortho}
		\langle \orthoterm(\xi), v \rangle_g = \omega_g(\gradient \phi_1(\xi), \dots, \gradient \phi_{n-1}(\xi), v), 
	\end{equation}
	where $\omega_g$ is the \textit{volume form}\footnote{The volume form exists since the manifold $\manifold$ is assumed to be oriented.} associated with $\manifold$ (its properties include skew-symmetry in its arguments) \cite[p. 30]{lee2018introduction}. The existence and uniqueness of $\gradient \phi_i(\xi)$ and $\orthoterm(\xi)$ are guaranteed by the Riesz representation theorem \cite[Theorem 6.42]{axler2015linear}. The calculations of these two terms are deferred until Section \ref{sec_calvf_manifold}. The guiding vector field defined on $\manifold$ is
	\begin{equation} \label{gvfm}
		\vf(\xi) = \orthoterm(\xi) - \sum_{i=1}^{n-1} k_i \phi_i(\xi) \gradient \phi_i(\xi). \tag{GVF-M}
	\end{equation}
	with $k_i>0$. In addition, as in the Euclidean case, the term $\orthoterm(\xi)$ is also orthogonal to each of the gradients $\gradient \phi_i(\xi)$, as formally stated in the following lemma:
	\begin{lemma}[Orthogonality] \label{lemma1}
	With definitions as above, there holds
	\[
			\langle \orthoterm(\xi), \gradient \phi_i(\xi) \rangle_g = 0
	\]
	for $i=1,\dots, n-1$ and $\xi \in \manifold$. 
	\end{lemma} 
	\begin{proof}
		This is an immediate consequence of the skew-symmetric property of the volume form $\omega_g$. 
	\end{proof}	
	
	We define $e: \manifold \to \mbr[n-1]$ by stacking $\phi_i$; that is,
	\begin{equation} \label{eqe}
	e(\xi) = \transpose{ (\phi_1(\xi) , \cdots , \phi_{n-1}(\xi))}.
	\end{equation}
	Using this notation, the desired path is equivalent to 
	\begin{equation} \label{eqpath2}
	\mathcal{P} = \{ \xi \in \manifold : e(\xi)=\bm{0} \}.
	\end{equation}
	This definition of the desired path suggests that $e(\xi)$ can be taken as the \emph{path-following error} between the point $\xi \in \manifold$ and the desired path $\set{P}$. The \emph{singular set} is defined by
	\begin{equation} \label{singularset}
	\begin{split}
		\set{C} &= \{ \xi \in \manifold : \vf(\xi)=\bm{0} \} \\
		&=\left\{ \xi \in \mathcal{M}: \orthoterm(\xi)=\sum_{i=1}^{n-1} k_i \phi_i(\xi) \gradient \phi_i(\xi)=\bm{0} \right\},
	\end{split}
	\end{equation}
	with the second equality following from Lemma \ref{lemma1}.	Note that the singular set $\set{C}$ may be empty or non-empty. To illustrate this, we provide two examples respectively below.
	
	\begin{example}[Non-empty $\set{C}$] \label{simple_example1}
		Consider that the desired path is a 2D unit circle in the Euclidean space $\mbr[2]$ described by $\phi(x,y)=x^2+y^2-1=0$; then one can obtain the vector field by \eqref{gvfe} and calculate that the singular set $\set{C}$ is a singleton consisting of the origin; that is, $\set{C} = \{(0,0)\}$.
	\end{example}
	\begin{example}[Empty $\set{C}$] \label{simple_example2}
		A simple example is a straight line in the Euclidean space $\mbr[2]$ described by $\phi(x,y)=y=0$, which is the $X$-axis. One can calculate that the propagation term is a non-zero constant vector (i.e., $E \nabla \phi = \transpose{(-1,0)}$, where  $E=\left[\begin{smallmatrix}0 & -1 \\ 1 & 0\end{smallmatrix}\right]$ is the $90^\circ$ rotation matrix). Due to the orthogonality of the propagation term and the convergence term of \eqref{gvfe}, this implies that the vector field $\vf(x,y) \ne \bm{0}$ in $\mbr[2]$, and thus the singular set $\set{C}$ is empty.
	\end{example}
	 Whether or not the singular set is empty may not be straightforward to determine, since one needs to obtain the analytic expression of the vector field and check if any point on the manifold renders it zero.
	 
	\begin{remark} \label{remark1}
		We emphasize the significance of studying the guiding vector field in \eqref{gvfm}. As the guiding vector field \eqref{gvfm} is a generalization of the particular case \eqref{gvfe} defined on the Euclidean space $\mbr[n]$, which is the space usually considered in the literature, it suffices to explain the importance (or generality) of this particular case \eqref{gvfe}. Firstly, different from many existing studies which restrict consideration to simple desired paths such as a circle or a straight line (or a combination of them)  \cite{Sujit2014,nelson2007vector,caharija2016integral}, the vector field in \eqref{gvfe} (or \eqref{gvfm}) is designed for any \emph{general} sufficiently smooth desired path. %
		Secondly, many vector fields in the literature are essentially \emph{variants} of the vector field in \eqref{gvfe}. For example, some variants are obtained by adding $\phi_i$-dependent gains to the convergence term or (and) the propagation term in \eqref{gvfe} \cite{michalek2018vfo,kapitanyuk2017guiding,liang2016combined}, and some by adding time-varying gains or an additional time-varying component \cite{Goncalves2010,lawrence2008lyapunov}. Thus, the vector fields in \cite{michalek2018vfo,kapitanyuk2017guiding,liang2016combined,yao2019integrated,yao2021collision} can be regarded as 2D specializations of \eqref{gvfe}, and those in \cite{lakomy2017,yao2018cdc,yao2020auto,kapitanyuk2017guiding2,liang2016combined} as 3D specializations of \eqref{gvfe}. Therefore, the study of the basic vector field in \eqref{gvfe} and its generalized counterpart \eqref{gvfm} is of great significance. In addition, to clearly observe the topological properties of the vector field, we do not consider time-varying components, and thus we focus on the \emph{autonomous} differential equation \eqref{eq1} with the vector field \eqref{gvfm}. %
	\end{remark}
	
	\subsection{Standing assumptions} \label{subsec_assump}
	In the remainder of the paper, the following assumptions will be made:
	\begin{assump} \label{assump1}
	There are no singular points on the desired path. More precisely, $\set{C}$ is empty or otherwise there holds $\dist(\mathcal{C}, \mathcal{P}) > 0$. 
	\end{assump}
	\begin{assump} \label{assump2}
	For any given constant $\kappa > 0$, there holds
	$
	\inf\{ ||e(\xi)||: \dist(\xi, \mathcal{P}) \ge \kappa\} > 0.
	$
	\end{assump}
	 Assumption \ref{assump1} ensures the ``regularity'' of the desired path $\set{P}$ stated in Lemma \ref{lemmanifold} below. 
	\begin{lemma}[Regularity of $\set{P}$] \label{lemmanifold}
		The zero vector $\bm{0} \in \mbr[n-1]$ is a regular value of the map $e$ in \eqref{eqe}, and hence the desired path $\mathcal{P}$ is a $C^2$ (properly) embedded submanifold in $\manifold$. 
	\end{lemma}	
	\begin{proof}
		This is a direct application of the regular level set theorem \cite[Corollary 5.14]{lee2015introduction}.
	\end{proof}

	Assumption \ref{assump2} implies that as the norm of the path-following error $\norm{e(\xi)}$ approaches zero, the trajectory $\xi(t)$ approaches the desired path $\set{P}$ \cite{yao2021dichotomy}. %
	These assumptions are vital in the sense that if either of these assumptions is not satisfied, then different choices of surface functions $\phi_i$ for the same desired path may lead to opposite convergence results, as the next example shows.
	
	\begin{example}[Opposite Convergence Results] \label{simple_example3}
	We consider a straight line in the 3D Euclidean space $\mbr[3]$. One can choose the surface functions $\phi_i,i=1,2$, as $\phi_1(x,y,z) = y, \phi_2(x,y,z) = z$, and the integral curves of the corresponding vector field \eqref{gvfe} converge to the desired straight line (see Fig. \ref{fig:ex2_1}). Another design of surface functions $\phi_i$ is $\phi_1(x,y,z) = y e^{-x}, \phi_2(x,y,z) = z$. In this case, however, as shown in Fig. \ref{fig:ex3_1}, the trajectory diverges from the desired path, although the norm of the path-following error $\norm{e}$ for this case is also approaching zero along the trajectory (see Fig. \ref{fig:ex_erros}). The reason is that the second case violates Assumption \ref{assump2}. This can be observed by considering a straight line $L$ parallel to the desired path but keeping a positive distance $\dist(L, \set{P})>0$. For example, let $L \defeq \{(x,1,0): x \in \mbr[]\}$. Then the fact that $\inf\{\norm{e(\xi)} : \xi \in L\} = 0$ violates Assumption \ref{assump2}.
	\begin{figure}[tb]
		\centering
		{
			\subfigure[]{
				\includegraphics[width=0.47\columnwidth]{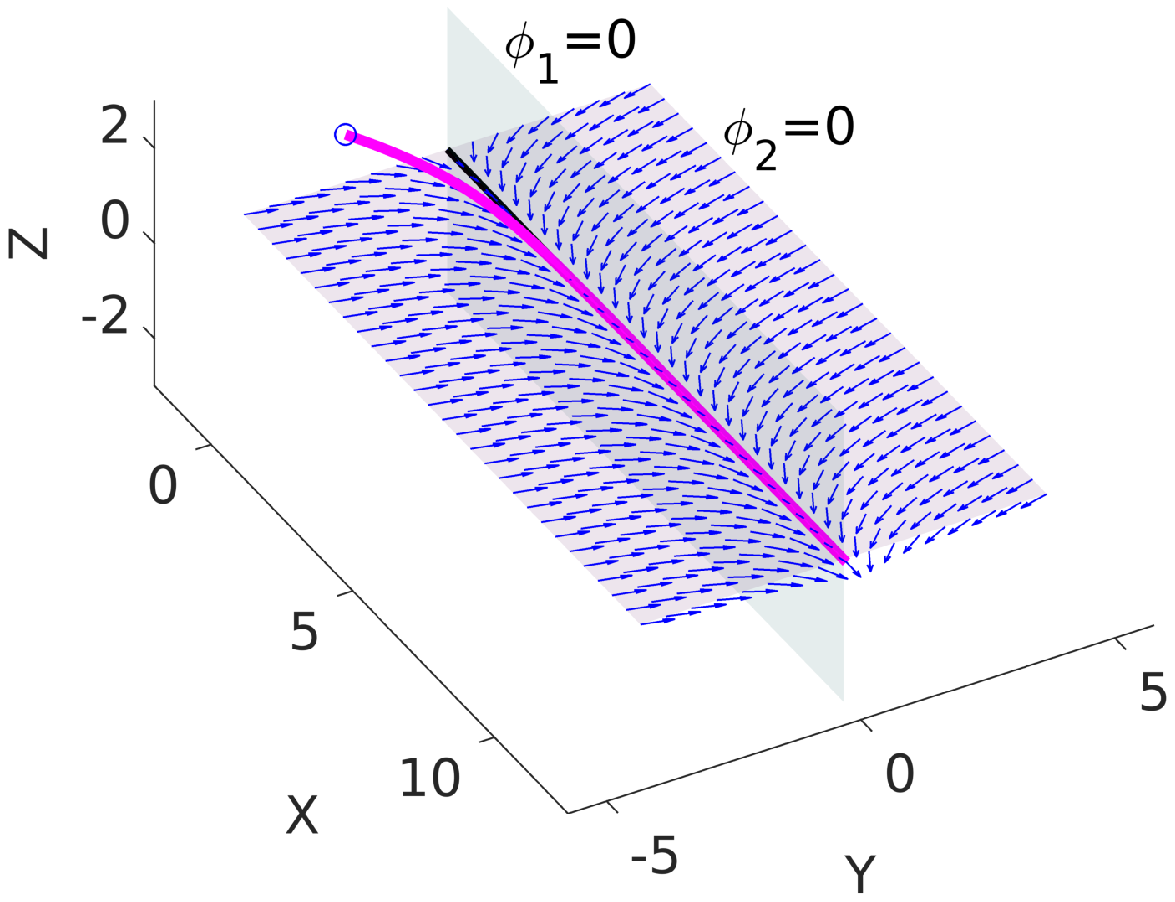}
				\label{fig:ex2_1}}%
			\subfigure[]{
				\includegraphics[width=0.5\columnwidth]{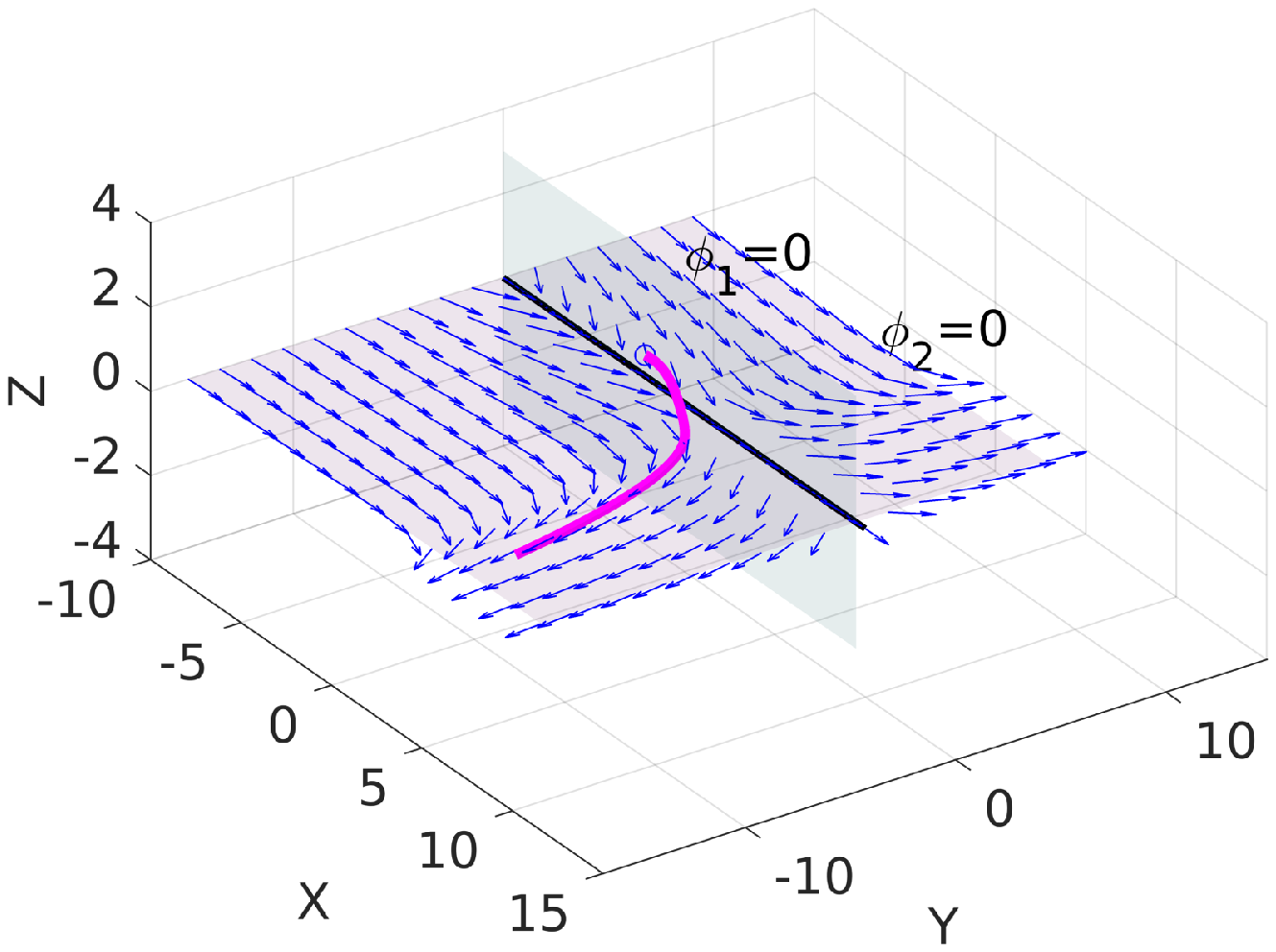}
				\label{fig:ex3_1}}\\
			\subfigure[]{
				\includegraphics[width=0.5\columnwidth]{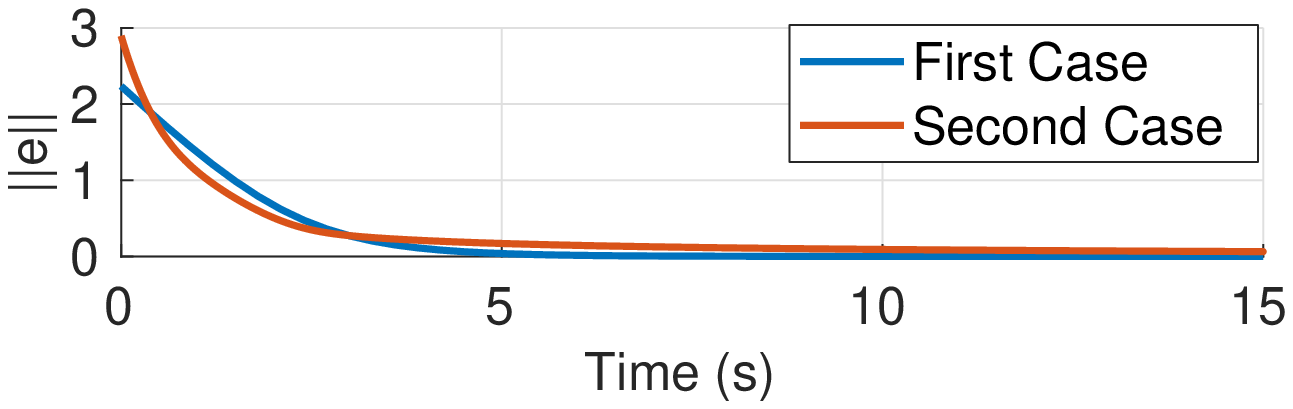}
				\label{fig:ex_erros}}%
		}
		\caption{The same desired path ($X$-axis) with different surface functions $\phi_i,i=1,2$. Magenta lines are trajectories, starting from the positions represented by blue points. \subref{fig:ex2_1} The trajectory converges to the desired path.  \subref{fig:ex3_1} The trajectory diverges from the desired path. \subref{fig:ex_erros} Both of the norms of the path-following error converge to zero.}
		\label{fig_ex_2}
	\end{figure}		 
	\end{example}

	\section{Computation of guiding vector fields on manifolds} \label{sec_calvf_manifold}
	The definitions of the gradient $\gradient \phi_i$ in $\eqref{eq_gradient}$ and the orthogonal term $\orthoterm$ in $\eqref{eq_ortho}$ for the guiding vector field on the manifold $\manifold$ are too abstract for direct computations. We present some general methods to compute these terms in coordinates. To this end, we suppose that the manifold $\manifold$ considered in \eqref{eqpath1} is an $n$-dimensional smooth submanifold embedded in the Euclidean space $\mbr[n+k]$, where $n$ is the dimension of the manifold $\manifold$ and $k$ is some positive integer\footnote{This is always possible according to the Whitney embedding theorem \cite[Theorem 6.15]{lee2015introduction}, which concludes that every smooth $n$-manifold admits a proper smooth embedding into $\mbr[2n+1]$.}, and the manifold $\manifold$ is a regular level set of a smooth function $F: \mbr[n+k] \to \mbr[k]$. Namely, we assume 
	\begin{equation} \label{eq_manifold_euclidean}
		\manifold = \inv{F}(a) = \{ x \in \mbr[n+k] : f_i(x)=a_i, i=1,\dots,k \},
	\end{equation}
	where $f_i: \mbr[n+k] \to \mbr[]$ are smooth component functions of $F$ and $a=(a_1,\dots,a_k) \in \mbr[k]$ is a regular value of $F$. For example, if the manifold $\manifold$ is the sphere $\mathbb{S}^2$, then it is a two-dimensional manifold embedded in $\mbr[3]$ and can be described by $\mathbb{S}^2 = \{x=(x_1,x_2,x_3) \in \mbr[3] : f_1(x)=x_1^2+x_2^2+x_3^2=1 \}$. %
	Similar examples can be found for some other common manifolds, such as $SO(3)$ and the torus $\mathbb{T}=\mathbb{S}^1 \times \mathbb{S}^1$. 
	
	For notational simplicity, let $m \defeq n+k$. In addition, to distinguish the Riemannian metrics in the Euclidean space $\mbr[m]$ and in the manifold $\manifold$, we adopt the following notations. For $x \in \mbr[m]$,  the Riemannian metric in the Euclidean space $\mbr[m]$ is the canonical one, denoted by $\langle \cdot, \cdot \rangle^{\mbr[m]}_x : T_x \mbr[m] \times T_x \mbr[m] \to \mbr[]$, while for $y \in \manifold$, the Riemannian metric is denoted by $\langle \cdot, \cdot \rangle^{\manifold}_y : T_y \manifold \times T_y \manifold \to \mbr[]$. 
	Let $\tilde{\phi_i}: \set{U} \to \mbr[]$, where $\set{U} \subseteq \mbr[m]$ is a neighborhood of $\manifold  \subseteq \mbr[m]$, be an extension of the surface function $\phi_i: \manifold \to \mbr[]$ in \eqref{eqpath1} \cite[Lemma 5.34]{lee2015introduction}; that is, the restriction of $\tilde{\phi_i}$ on $\manifold$ is $\phi_i$, or $\tilde{\phi_i} \big|_{\manifold} = \phi_i$. %
	
	The following result shows that the gradient defined on the manifold $\manifold$ is just the orthogonal projection of the ``usual'' gradient in the Euclidean space onto the tangent space to the manifold $\manifold$ at some point.
	\begin{prop} \label{prop2}
		For $x \in \manifold \subseteq \mbr[m]$, define the orthogonal projection function $\text{Pr}_{T_x \manifold}: T_x \mbr[m] \to T_x \manifold$. Then we have
		\begin{equation} \label{eq_gradphi1}
			\gradient \phi_i (x) = \text{Pr}_{T_x \manifold} (\nabla \tilde{\phi_i} (x)),
		\end{equation}
		where $\gradient \phi_i (x) \in T_x \manifold$ and $\nabla \tilde{\phi_i} (x) \in T_x \mbr[m]$. In particular,
		\begin{equation} \label{eq_gradphi2}
		\scalemath{0.9}{
			\gradient \phi_i (x) = \nabla \tilde{\phi_i} (x) - \sum_{j=1}^k \frac{\langle \nabla \tilde{\phi_i} (x), \nabla f_j (x) \rangle^{\mbr[m]}_x}{\| \nabla f_j (x) \|^2} \nabla f_j (x),
			}
		\end{equation}
		where $f_j, j=1,\dots,k,$ are functions in \eqref{eq_manifold_euclidean}.
	\end{prop}
	\begin{proof}
		The equation \eqref{eq_gradphi1} is a standard result \cite[pp. 360-362]{helmke2012optimization}. Since the manifold $\manifold$ is described by \eqref{eq_manifold_euclidean}, we have that $(T_x \manifold)^{\bot} = \text{span}( \nabla f_1 (x), \dots, \nabla f_k (x))$, hence \eqref{eq_gradphi2}.
	\end{proof}

	Using these computable gradients $\gradient \phi_i, \, i=1,\dots, n-1,$ as in Proposition \ref{prop2}, we can now derive a computable form for the orthogonal term $\orthoterm$ in \eqref{eq_ortho}. Before that, recall that for $p_i=\transpose{(p_{i1}, p_{i2}, p_{i3})} \in \mathbb{R}^3$, $i=1,2$, the cross product $p_1 \times p_2$ is calculated by the following intuitive \emph{formal expression} involving the matrix determinant \cite[pp. 241-242]{fraleigh1995linear}:%
	\begin{multline*} 
	p_1 \times p_2 = \vmatr{ \bm{b_1} & \bm{b_2} & \bm{b_3} \\ p_{11} & p_{12} & p_{13} \\ p_{21} & p_{22} & p_{23}}  
	= \vmatr{p_{12} & p_{13} \\ p_{22} & p_{23}} \bm{b_1} - \\ \vmatr{p_{11} & p_{13} \\ p_{21} & p_{23} } \bm{b_2}  + \vmatr{p_{11} & p_{12} \\ p_{21} & p_{22} } \bm{b_3}  
	= \vmatr{ p_{11} & p_{21} &  \bm{b_1} \\ p_{12} & p_{22} & \bm{b_2} \\ p_{13} & p_{23} & \bm{b_3}},  
	\end{multline*}
	where $\bm{b_j} \in \mathbb{R}^3, \, j=1,2,3,$ are the standard basis vectors and $|\cdot|$ is the determinant of a square matrix. The second equality is obtained using the cofactor expansion along the first row of the matrix in the first equality, where in evaluating the determinant, the $\bm{b_i}$ should initially be regarded as scalars, and in the final evaluation replaced by the basis vectors which they are. This formal expression can be naturally extended to Euclidean spaces of any dimension, a fact utilized below.
	\begin{prop} \label{prop3}
		For $x \in \manifold \subseteq \mbr[m]$, the orthogonal term $\orthoterm$ defined in \eqref{eq_ortho} can be computed by the following formal form:
		\begin{multline} \label{eq_ortho_cal}
		\orthoterm (x) = \\
		\scalemath{0.9}{
			\det \left[ \nabla f_1 (x), \cdots , \nabla f_k (x), \gradient \phi_1 (x), \cdots,  \gradient \phi_{n-1} (x),  \begin{matrix} \bm{b_1} \\ \vdots \\ \bm{b_{m}} \end{matrix}  \right], 
			}
		\end{multline}
		where $\bm{b_i} \in \mbr[m], i=1,\dots,m,$ are standard basis vectors, and $\gradient \phi_i, \, i=1,\dots, n-1,$ are calculated from Proposition \ref{prop2}.
	\end{prop}
	\begin{proof}
		We first consider the case for the Euclidean space $\mbr[m]$, and then extend to that for the manifold $\manifold$. For any $x \in \mbr[m]$, we can pick a volume form $\omega_x: \underbrace{T_x \mbr[m]\times \dots \times T_x \mbr[m]}_{m} \to \mbr[]$, which is a skew-symmetric and non-degenerate linear function, such that $\omega_x$ is smooth with respect to $x$. Note that the general form of $\omega_x$ is $\omega_x = c(x) \cdot d x_1 \wedge \dots \wedge d x_{m}$, where $c: \mbr[m] \to \mbr[]$ is a nonzero and smooth function. Specifically, for the column vectors $a_i=\transpose{(a_{i,1},\dots,a_{i,m})} \in \mbr[m], \, i=1,\dots,m$, it holds that 
		\begin{equation}
			d x_1 \wedge \dots \wedge d x_{m} \,(a_1,\dots, a_{m})  = 
				\det \matr{a_{1,1} & \cdots & a_{1,m} \\ \vdots & \ddots & \vdots \\ a_{m,1} & \cdots & a_{m,m}}.
		\end{equation}
		Let 
		$
			\omega_x \big( a_1, \dots, a_{m}\big) \defeq d x_1 \wedge \dots \wedge d x_{m} \,(a_1, \dots, a_{m}).
		$
		We can calculate the volume form on the manifold $\manifold$, denoted by $\omega^\manifold: \underbrace{T_x \manifold \times \dots \times T_x \manifold}_{n} \to \mbr[]$, as follows:
		\begin{align*}
			\omega_x^\manifold (v_1, \dots, v_n) \defeq& 
			\omega_x \left( \nabla f_1 (x)  , \cdots , \nabla f_k (x), v_1, \dots, v_n \right) \\
			=& \det \left[ \nabla f_1 (x), \cdots, \nabla f_k (x),  v_1, \dots, v_n \right],
		\end{align*}
		where $v_j \in T_x \manifold, j=1,\dots,n$. Hence, by \eqref{eq_ortho}, we have
		$
			\langle \orthoterm(\xi), \cdot \rangle^\manifold_x = \omega^\manifold_x (\gradient \phi_1(x), \dots, \gradient \phi_{n-1}(x), \cdot) 
			= 	\langle \triangle, \, \cdot \rangle^\manifold_x,
		$
		where $\triangle$ is the right-hand side of \eqref{eq_ortho_cal}. %
	\end{proof}

	The above results serve collectively as a general procedure to compute the guiding vector field on $\manifold$.

	 \begin{figure}[tb]
	\centering
	\includegraphics[width=0.4\columnwidth]{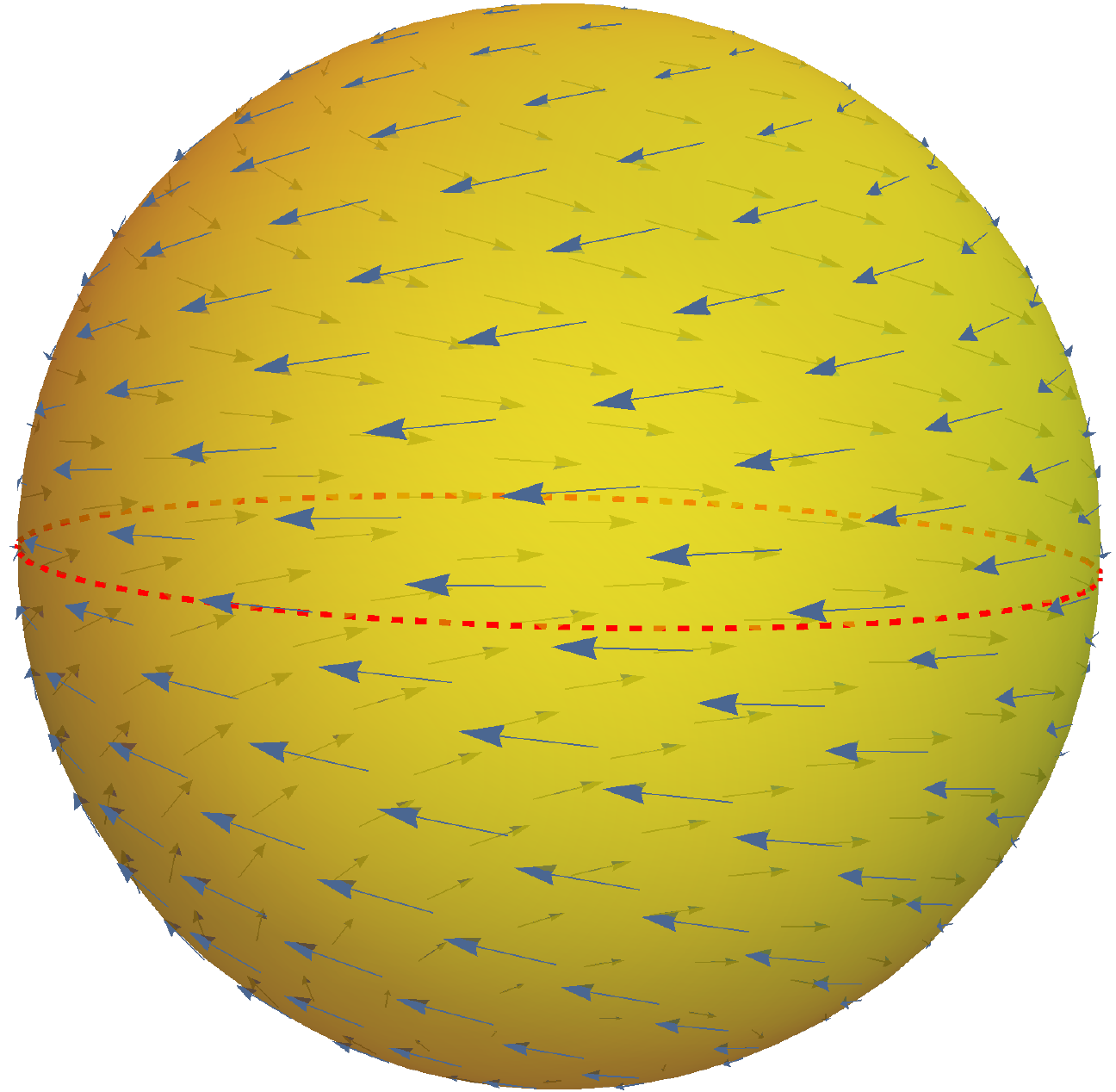}
	\caption{The guiding vector field $\vf$ on the sphere $\manifold=\mathbb{S}^2$ in Example \ref{ex4}. The red dashed line is the desired path $\set{P}$, which is a circle on the sphere.}
	\label{fig:spherevf}
	\end{figure}
	\begin{example}[Guiding vector field on $\mathbb{S}^2$] \label{ex4}
		Suppose the sphere $\mathbb{S}^2$ is the manifold $\manifold$ on which the guiding vector field is defined. It is a two-dimensional manifold that can be naturally embedded in the Euclidean space $\mbr[3]$, and then \eqref{eq_manifold_euclidean} becomes
		\[
		\manifold = \mathbb{S}^2 = \{ (x,y,z) \in \mbr[3] : f_1(x,y,z)=x^2+y^2+z^2=1 \}.
		\]
		The desired path $\set{P} \subseteq \manifold$ is the unit circle with $z=0$. Therefore, we can let $\tilde{\phi}_1: \mbr[3] \to \mbr[]$ be defined by $\tilde{\phi}_1(x,y,z)=z$, and then $\phi_1: \manifold \to \mbr[]$ is just $\phi_1 = \tilde{\phi}_1 \big|_{\manifold}$ in \eqref{eqpath1}. We can calculate the gradient term $\gradient \phi_1$  and the orthogonal term $\orthoterm$ by Proposition \ref{prop2} and Proposition \ref{prop3} respectively. First, we have $\nabla \tilde{\phi}_1=\transpose{(0,0,1)}$ and $\nabla f_1 = \transpose{(2x, 2y, 2z)}$. For any point $\xi \in \manifold$, it follows from \eqref{eq_gradphi2} that
		\[
		\gradient \phi_1(\xi) = \nabla \tilde{\phi}_1(\xi) - \frac{\langle \nabla \tilde{\phi}_1(\xi), \nabla f_1(\xi) \rangle}{\norm{\nabla f_1(\xi)}^2} \nabla f_1(\xi) = \matr{-xz \\ -yz \\ 1-z^2},
		\]
		and from \eqref{eq_ortho_cal} that
		\[
		\orthoterm(\xi) = \det \left[ \nabla f_1(\xi), \, \gradient \phi_1(\xi),\, \begin{matrix} \bm{b_1} \\ \bm{b_2} \\ \bm{b_3} \end{matrix} \right] = \matr{2y \\ -2x \\ 0}.
		\]	
		Finally, the ``computable'' guiding vector field on the sphere is obtained by putting these two terms into \eqref{gvfm} as follows:
		$
		\vf(\xi) = \orthoterm(\xi) -  k_1 \phi_1(\xi) \gradient \phi_1(\xi) = \big( 2y + k_1 x z^2 , -2x + k_1 y z^2 , k_1 z (z^2-1) \big)^\top
		$
		for any point $\xi=(x,y,z) \in \mathbb{S}^2$. The vector field is shown in Fig. \ref{fig:spherevf}. Interestingly, there are two singular points in this vector field: the north pole and the south pole (i.e., $(0,0,\pm 1)$). However, as shown in Example \ref{simple_example1}, if we consider the Euclidean space $\manifold=\mbr[2]$, then there is only one singular point: the origin $(0,0)$. Similarly, if we consider the 3D Euclidean space $\manifold=\mbr[3]$, and use the functions $\phi_1(x,y,z)=x^2+y^2-1=0$ and $\phi_2(x,y,z)=z=0$ to characterize the same unit circle as the desired path, there is now only one singular point and it is at the origin $(0,0,0)$. This example shows that corresponding to the same desired path, guiding vector fields (and singular sets) defined on different manifolds are possibly quite different.		
 	\end{example}

	\section{Dichotomy convergence and stability analysis} \label{sec_dichotomy}
	It is important to analyze the convergence results of the integral curves of the vector field \eqref{gvfm}; that is, the trajectories of the differential equation \eqref{eq1}, where the vector field $\vf(\cdot)$ is defined in \eqref{gvfm}. %
	It turns out that the dichotomy convergence property holds not only for the vector field defined on the Euclidean space $\mbr[n], n \ge 2$ \cite{yao2018cdc}, but also for that on the Riemannian manifold $\manifold$. First we define the function $V: \manifold \to \mbr[]$ as
	\begin{equation} \label{eqlyapunov}
		V(\xi)=\transpose{e}(\xi) K e(\xi),
	\end{equation}
	where $K \defeq \diag{k_1, \dots, k_{n-1}}$ is the diagonal matrix with all the positive gains $k_i$, $i=1,\dots,n-1$. The function $V$ is non-negative and attains zero if and only if $\xi \in \set{P}$. This function is utilized as a Lyapunov-like function in the analysis subsequently. We assume until further notice that the desired path $\set{P}$ is homeomorphic to the unit circle (hence compact):
	\begin{assump} \label{assump4}
		The desired path $\set{P}$ is homeomorphic to the unit circle $\mathbb{S}^1$ (i.e., $\set{P} \approx \mathbb{S}^1$). 
	\end{assump}
	We can choose $r>0$ sufficiently large such that the open ball $\set{B}_r \defeq \{x \in \manifold : \norm{x} < r \}$ contains the desired path $\set{P}$ (i.e., $\set{P} \in \interior \set{B}_r$). Let $\alpha' \defeq \min_{p \in \partial \set{B}_r} V(p)>0$, where the minimum is attained on the compact sphere $\partial \set{B}_r$ (i.e., the boundary of the ball $\set{B}_r$), and it is positive since $\partial \set{B}_r \cap \set{P} = \emptyset$. We can choose a positive constant $\alpha$ such that $0<\alpha<\alpha'$, and the set $\Omega_\alpha$ defined below is compact:
	\begin{equation} \label{eq_omega_alpha}
			\Omega_\alpha \defeq \{\xi \in \set{B}_r : V(\xi) \le \alpha \}.
	\end{equation}
	Note that $\set{P} \subseteq \Omega_\alpha$ for any $\alpha>0$. Now we can present the dichotomy convergence result as follows:
	\begin{theorem}[Dichotomy convergence] \label{thm_dichotomy}
		Consider the autonomous system \eqref{eq1}, where the vector field $\vf: \manifold \to T \manifold$ is in \eqref{gvfm}. Then the compact set $\Omega_\alpha$ in \eqref{eq_omega_alpha} is positively invariant. In addition, %
		every trajectory of \eqref{eq1} starting from $\Omega_\alpha$ converges to either the desired path $\set{P}$, or the singular set $\set{C}$ as $t \to \infty$ (i.e., the dichotomy convergence property holds).
	\end{theorem}
	\begin{proof}
		The proof for a special case $\manifold=\mbr[3]$ (but easily generalizable to $\manifold=\mbr[n]$) is presented in \cite{yao2018cdc}. It can be further generalized to the case of a Riemannian manifold $\manifold$ by modifying the involved calculations related to the Riemannian metric. First note that
		\begin{equation} \label{eq_ortho_zero}
		\scalemath{0.9}{
			\begin{split}
				\langle \gradient \phi_i, \vf \rangle_g \stackrel{\eqref{gvfm}}{=} 
				&\left\langle \gradient \phi_i, \; \orthoterm(\xi) - \sum_{j=1}^{n-1} k_j \phi_j(\xi) \gradient \phi_j(\xi) \right\rangle_g \\
				= & \left\langle \gradient \phi_i, - \sum_{j=1}^{n-1} k_j \phi_j(\xi) \gradient \phi_j(\xi) \right\rangle_g,
			\end{split}
		}
		\end{equation}
		for $i=1,\dots,n-1$, where we have used the orthogonality property (Lemma \ref{lemma1}) in the last equation. Now we can calculate the time derivative of the path-following error $e$:
		\begin{equation} \label{eq_e_dot}
		\scalemath{0.9}{
			\begin{split} 
			\dt e(\xi(t)) &= \dt \matr{\phi_1(\xi(t)) \\ \vdots \\ \phi_{n-1}(\xi(t))} = \matr{\langle \gradient \phi_1, \vf \rangle_g \\ \vdots \\ \langle \gradient \phi_{n-1}, \vf \rangle_g} \\
			&\stackrel{\eqref{eq_ortho_zero}}{=} \matr{\langle \gradient \phi_1, -\sum_{j=1}^{n-1} k_j \phi_j \gradient \phi_j \rangle_g \\ \vdots \\ \langle \gradient \phi_{n-1}, -\sum_{j=1}^{n-1} k_j \phi_j \gradient \phi_j \rangle_g}.
			\end{split}
		}
		\end{equation}
		Therefore, the time derivative of the Lyapunov function \eqref{eqlyapunov} is
		\begin{equation} \label{eq_vdot_manifold}
		\scalemath{0.9}{
			\begin{split}
			\dt V &= 2 \transpose{\left( \dt e \right) } K e \\
			&\stackrel{\eqref{eq_e_dot}}{=} 2 \transpose{ \matr{\langle \gradient \phi_1, -\sum_{j=1}^{n-1} k_j \phi_j \gradient \phi_j \rangle_g \\ \vdots \\ \langle \gradient \phi_{n-1}, -\sum_{j=1}^{n-1} k_j \phi_j \gradient \phi_j \rangle_g}} \matr{k_1 \phi_1 \\ \vdots \\ k_{n-1} \phi_{n-1}} \\
			&= -2 \left\langle \sum_{j=1}^{n-1} k_j \phi_j \gradient \phi_j, \sum_{j=1}^{n-1} k_j \phi_j \gradient \phi_j \right\rangle_g \le 0.
			\end{split}
		}
		\end{equation}
		Due to the negative semi-definiteness of \eqref{eq_vdot_manifold}, the compact set  $\Omega_\alpha$ is positively invariant. Next we use the the LaSalle's invariance principle \cite[Theorem 4.4]{khalil2002nonlinear} to conclude the convergence results. First, we have the following equivalent sets
		$
			\set{I} \defeq \left\{\xi \in \manifold: \dt V(\xi) = 0 \right\} = 
			\left\{\xi \in \manifold: \sum_{j=1}^{n-1} k_j \phi_j(\xi) \gradient \phi_j(\xi) = 0 \right\} = \set{P} \cup \set{C}.
		$
		The last equality is justified as follows. If a point $p \in \set{P} \cup \set{C}$, then it is obvious that $p$ is contained in the set on the left-hand side of the equality, and thus $\set{P} \cup \set{C}$ is a subset of the set on the left-hand side. If a point $p$ is in the set of the left-hand side, then we have $\sum_{j=1}^{n-1} k_j \phi_j(p) \gradient \phi_j(p) = 0$. This implies that either all gradients $\gradient \phi_j(p)$ are linearly independent and $\phi_j(p) = 0$ for all $j=1,\dots,n-1$, or the gradients $\gradient \phi_j(p)$ are linearly dependent and hence $\orthoterm(p)=0$. In the former case, $p \in \set{P}$, and in the latter case, $p \in \set{C}$. Therefore, the set on the left-hand side is a subset of $\set{P} \cup \set{C}$. Combining these two arguments, the last equality holds. 
		
		It is easy to see that the largest invariant set in $\set{I} \cap \Omega_{\alpha}$ is itself: every trajectory of \eqref{eq1} starting from $\set{C} \cap \Omega_{\alpha}$ will remain in $\set{C} \cap \Omega_{\alpha}$ because $\set{C} \cap \Omega_{\alpha}$ consists of equilibrium points of \eqref{eq1}, and every trajectory of \eqref{eq1} starting from $\set{P} \cap \Omega_{\alpha}=\set{P}$ will remain in $\set{P}$ because the guiding vector field in \eqref{gvfm} degenerates to $\vf(\xi) = \orthoterm(\xi)$ on $\set{P}$. Therefore, according to the LaSalle's invariance principle \cite[Theorem 4.4]{khalil2002nonlinear}, all trajectories starting from the compact and positively invariant set $\Omega_\alpha$ will converge to the largest invariant set $\set{I} \cap \Omega_{\alpha}$. By Assumption \ref{assump1}, this implies that trajectories either converge to the desired path $\set{P}$ or the singular set $\set{C}$, hence the dichotomy convergence property still holds.
	\end{proof}
	
	We have shown by Theorem \ref{thm_dichotomy} that some undesirable phenomena in nonlinear systems, such as chaos, finite-time escape, cannot occur for the system \eqref{eq1} with the vector field \eqref{gvfm}. Now we show another desirable property: the desired path $\set{P}$ is attractive, while the singular set $\set{C}$ is not.

	\begin{defn} \label{def1} %
		A non-empty closed positively invariant set $\set{A} \subseteq \manifold$ is \emph{attractive} with respect to \eqref{eq1}, if there exists an open neighborhood $\set{U}$ of $\set{A}$ such that every trajectory $\xi(t)$ of \eqref{eq1} that starts within $\set{U}$ (i.e., $\xi(0) \in \set{U}$) converges (topologically) to $\set{A}$ in the sense that for any neighborhood $\set{V}$ of $\set{A}$, there exists a $T>0$, such that $\xi(t \ge T) \subseteq \set{V}$ when $\xi(0) \in \set{U}$ (this implies that $\dist(\xi(t), \set{A}) \to 0$ as $t \to \infty$). If the set $\set{A}$ is not attractive, then it is called \textit{non-attractive}. The set of all points for which trajectories start from and converge (topologically) to $\set{A}$ is the domain of attraction of $\set{A}$ (obviously, $\set{U}$ is a subset of the domain of attraction of $\set{A}$).
	\end{defn}
	
	Note that in the definition above, $\set{A}$ is not required to be compact. Note also that a set can be non-attractive and yet trajectories from outside the set can converge to the set; consider for example a time-invariant linear system $\dot x=A x$, where some eigenvalues of $A$ are in the left half plane and some are in the right half plane, resulting in the origin being non-attractive. We can also define the (Lyapunov) stability and asymptotic stability of the set $\set{A}$ \cite[Definition 4.10]{haddad2011nonlinear} as below.
	\begin{defn} \label{def2}
		A non-empty closed positively invariant set $\set{A}$ is \emph{(Lyapunov) stable} with respect to \eqref{eq1}, if for every open neighborhood $\set{U}_1$ of $\set{A}$, there exists an open neighborhood $\set{U}_2 \subseteq \set{U}_1$ of $\set{A}$, such that every trajectory of \eqref{eq1} stays in $\set{U}_1$ once it starts from $\set{U}_2$ (i.e., $\xi(t) \in \set{U}_1$ for $t \ge 0$ with $\xi(0) \in \set{U}_2$). Furthermore, if $\set{A}$ is both Lyapunov stable and attractive, then it is called \emph{asymptotically stable}. %
	\end{defn}

	\begin{coroll}[Asymptotic stability of $\set{P}$] \label{coroll_attra_P}
		The desired path $\set{P}$ is asymptotically stable.
	\end{coroll}
	\begin{proof} 
	Due to Assumptions \ref{assump1} and \ref{assump2}, there always exists a sufficiently small positive constant $\alpha$ such that $\Omega_\alpha \cap \set{C} = \emptyset$. Therefore, by Theorem \ref{thm_dichotomy}, the desired path $\set{P}$ is attractive. To prove that $\set{P}$ is asymptotically stable, we need to additionally show that it is (Lyapunov) stable. Define the set $\Gamma_a \defeq \{ p \in \manifold : \norm{e(p)} < a \}$ for some positive constant $a>0$, and it is obvious that $\set{P} \subseteq \Gamma_a$. By the Lyapunov argument in \eqref{eq_vdot_manifold} and Theorem 4.8 in \cite{khalil2002nonlinear}, the equilibrium point $e=0$ of the non-autonomous system \eqref{eq_e_dot} is uniformly stable. That is, for any $\epsilon>0$, there is $\delta>0$ (independent of the initial time instant $t_0$), such that $\xi(t_0) \in \Gamma_\delta \implies \xi(t) \in \Gamma_\epsilon$ for all $t \ge t_0 \ge 0$. For any open neighborhood $\set{U}_1$ of $\set{P}$, we can choose a positive constant $\epsilon$ sufficiently small\footnote{The existence of $\epsilon$ is guaranteed by the compactness of $\set{P}$ and Assumption \ref{assump2}. In fact, Assumption \ref{assump2} can be dropped, but the set $\Gamma_\epsilon$ should be changed to its component (i.e., the maximal connected subset of $\Gamma_\epsilon$) that contains $\set{P}$, and similarly, the set $\Gamma_\delta$ used in the subsequent part of the proof should also be changed to its component that contains $\set{P}$. }, such that $\Gamma_\epsilon$ is contained in $\set{U}_1$ (i.e., $\Gamma_\epsilon \subseteq \set{U}_1$). Due to the uniform stability of $e=0$, there exists $0<\delta<\epsilon$, such that $\xi(t) \in \Gamma_\epsilon$ for $t \ge 0$ whenever $\xi(0) \in \Gamma_\delta$. By letting $\set{U}_2 = \Gamma_\delta$ in Definition \ref{def2}, $\set{P}$ is asymptotically stable.
	\end{proof}

	\begin{lemma} \label{lem_nonattra_C}
	Suppose every trajectory starting at any point in $\manifold$ converges to either the desired path $\set{P}$ or the singular set $\set{C}$ as $t \to \infty$. Then the desired path $\set{P}$ and the singular set $\set{C}$ cannot be both attractive.
	\end{lemma}
	
	\begin{proof}		 %
	If $\set{C}=\emptyset$, then $\set{C}$ is non-attractive, and the claim is vacuously true. Thus we assume that $\set{C} \ne \emptyset$.  We first show that the domain of attraction of a non-empty closed attractive set $\set{A}$ of a dynamical system is open\footnote{This result is similar to Proposition 4.15 in \cite[Chapter V]{bhatia2002stability}, but the latter does not provide a proof.}, as a generalization of the standard result where this attractive set is replaced by an equilibrium point \cite[Proposition 5.44]{sastry2013nonlinear}. Since $\set{A}$ is attractive, by Definition \ref{def1}, there exists some open neighborhood $\set{U}$ of $\set{A}$ such that for any $x \in \set{U}$, the trajectory starting from $x$ converges (topologically) to $\set{A}$. Therefore, for any point $y$ in the domain of attraction of $\set{A}$, there exists a time $T>0$ such that $\Psi(T, y) \in \set{U}$ (because we choose $\set{V}=\set{U}$ in Definition \ref{def2}), where $\Psi: \mbr[]_{\ge0} \times \manifold \to \manifold$ denotes the flow of the dynamical system. By the continuity of $\Psi(T, \cdot)$ with respect to the second argument, there is some open neighborhood $\set{B}$ of $y$ such that $\Psi(T, \set{B}) \subseteq \set{U}$. Therefore, for all points $b \in \set{B}$, any trajectory starting from $b$ will go through $\Psi(T, b) \in \set{U}$ and converge to $\set{A}$ (by the existence and uniqueness of trajectory \cite[Theorem 3.1]{khalil2002nonlinear}), implying that $\set{B}$ is an open subset of the domain of attraction of $\set{A}$. Therefore, the domain of attraction of the attractive set $\set{A}$ is indeed open.
	
	We now argue by contradiction by assuming that both $\set{C}$ and $\set{P}$ are attractive. Thus, their domains of attraction are both open. Therefore, the whole configuration space $\manifold$ consists of only two kinds of points, those converging to $\set{C}$ and those to $\set{P}$ by the global dichotomy convergence property in Theorem \ref{thm_dichotomy}. This means that the configuration space $\manifold$ is a union of two disjoint open subsets, which is not possible since $\manifold$ is assumed to be connected.  
	\end{proof}		
	\begin{coroll}[Non-attractiveness of $\set{C}$]  \label{cor_nonattra_C}
	Under the hypotheses of Lemma \ref{lem_nonattra_C},  the singular set $\set{C}$ is non-attractive.		
	\end{coroll}
	\begin{proof}
		By Corollary \ref{coroll_attra_P}, the desired path $\set{P}$ is attractive. Under
		the hypotheses of Lemma \ref{lem_nonattra_C}, the desired path $\set{P}$ and the singular set $\set{C}$ cannot be both attractive. Therefore, the singular set $\set{C}$ is non-attractive.	
	\end{proof}
	\begin{remark} \label{remark3}
		Corollary \ref{cor_nonattra_C} is equivalent to saying that if Theorem \ref{thm_dichotomy} holds globally (i.e., $\Omega_{\alpha}$ can be replaced by $\manifold$), then the singular set $\set{C}$ is non-attractive. Theorem \ref{thm_dichotomy} holds globally, if $\set{M}$ is compact, or if $e(\xi)$ is radially unbounded (i.e., $\norm{e(\xi)} \to \infty$ as $\norm{\xi} \to \infty$). The radial unboundedness of  $e(\xi)$ is consistent with physical intuition, and is probably not restrictive in practice (e.g., it is true for Example \ref{simple_example1}, a typical case in the literature and in practice). Note that Theorem \ref{thm_dichotomy} does hold globally for all examples where the desired path is compact in this paper (i.e., Example \ref{simple_example1}, \ref{ex4}, \ref{ex5} and all examples in Section \ref{sec_simulation}).
	\end{remark}
	\begin{remark}
		If the singular set $\set{C}$ is non-attractive, by Definition \ref{def1}, it is still possible that some trajectories (commencing outside $\set{C}$) can converge to the singular set $\set{C}$. Nevertheless, in this case, one can immediately conclude that there must be some other trajectories that do not converge to $\set{C}$, no matter how near they start to the singular set $\set{C}$. 
	\end{remark}
	Note that by Theorem \ref{thm_dichotomy}, Corollary \ref{coroll_attra_P} and Corollary \ref{cor_nonattra_C}, we  \emph{cannot} conclude that trajectories converge to the desired path from \emph{almost all }initial conditions (i.e., \emph{almost global} convergence to $\set{P}$). The claim about almost global convergence to the desired path can be refuted simply by an example below where the singular set is of measure \emph{non-zero}. 
	\begin{example}[$\set{C}$ of measure \emph{non-zero}] \label{ex5}
		If the singular set is of measure non-zero, one cannot expect \emph{almost global }convergence to the desired path since every trajectory starting from the singular set will remain in that set. To construct such a case, first we introduce a smooth but non-real-analytic function (see Fig. \ref{subfig: non-analytic-func}) $b: \mbr[2] \to \mbr[]$:
		\begin{equation} \label{bump_func}
		\scalemath{0.9}{
		b(x,y) = \begin{dcases}
		\expo{\frac{1}{1-x^2-y^2}} & \text{if } x^2+y^2>1, \\
		0 & \text{otherwise}.
		\end{dcases}
		}
		\end{equation}
		We can construct the function $\phi: \mbr[2] \to \mbr[]$ using \eqref{bump_func} as below:
		\begin{equation} \label{non_analytic_phi}
			\phi(x,y) = 4 + (-x^2-y^2) \cdot b(x,y).
		\end{equation}
		The desired path $\set{P} = \{(x,y) \in \mbr[2]: \phi(x,y)=0 \}$ is a circle of radius approximately $2$. Moreover, the singular set $\set{C}$ of the vector field derived from $\phi$ is a disk of radius $1$ centered at the origin (see Fig. \ref{subfig: non-analtic-vf}). In this case, almost global convergence to the desired path is not possible, since the singular set $\set{C}$ has a non-zero measure.
		\begin{figure}[tb]
			\centering
			\subfigure[]{
				\includegraphics[width=0.4\columnwidth]{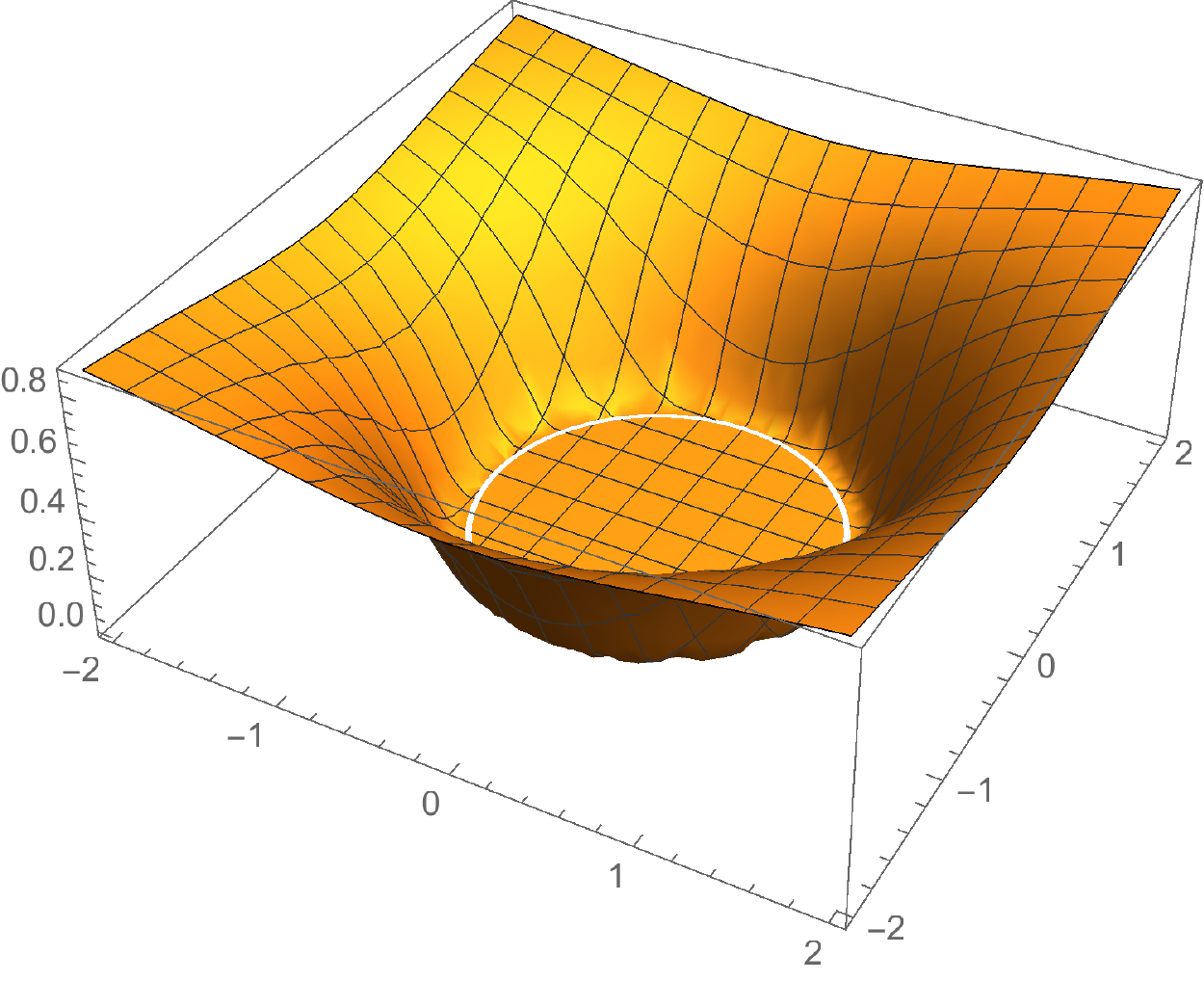}
				\label{subfig: non-analytic-func}
			}%
			\subfigure[]{
				\includegraphics[width=0.4\columnwidth]{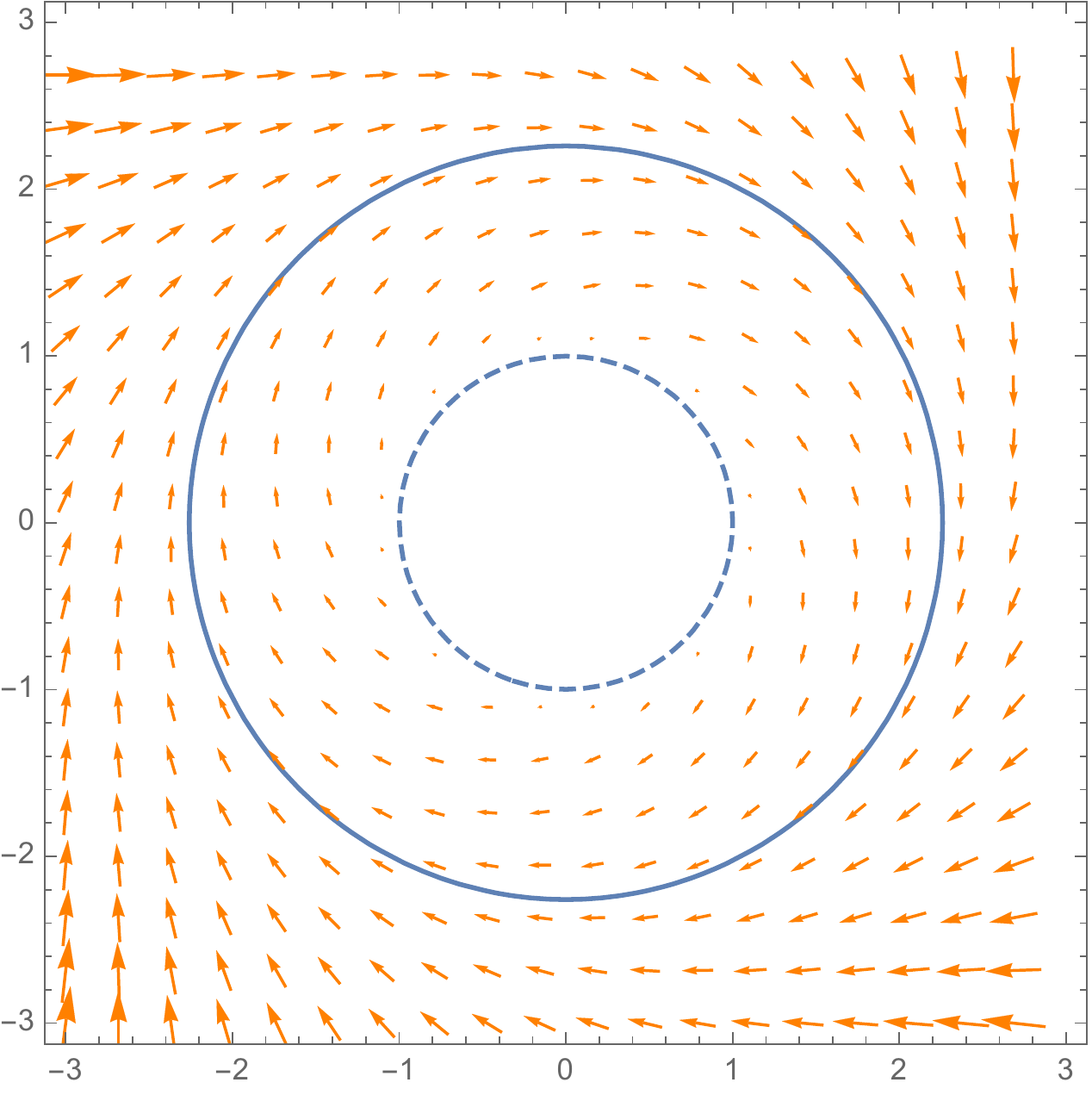}
				\label{subfig: non-analtic-vf}
			}
			\caption{Example \ref{ex5}. \subref{subfig: non-analytic-func} The graph of the smooth non-real-analytic function $b$ in \eqref{bump_func}. \subref{subfig: non-analtic-vf} The non-real-analytic vector field of which the function $\phi$ is given in \eqref{non_analytic_phi}. The solid line is the desired path and the dashed line is the boundary of the singular set, which is the unit disk centered at the origin. }
			\label{fig:non_analytic_vf}
		\end{figure}
	\end{example}
	Note that the function $\phi$ in the example above is \emph{not} real-analytic. In fact,  it is proved in \cite{Goncalves2010} that when all the surface functions $\phi_i$ are real analytic, then the singular set of the corresponding vector field is of measure zero. However, even if the surface functions $\phi_i$ are real analytic, we still cannot conclude that the initial conditions such that trajectories converging to the singular set are of measure zero, and thus almost global convergence to the desired path cannot be guaranteed. A well-known example where an equilibrium, which is obviously of measure zero, is \emph{non-attractive} but almost all trajectories converge to it is presented in \cite[p. 186]{sastry2013nonlinear}.

	The discussion above shows that it is generally challenging to study the domain of attraction of the desired path $\set{P}$ or the singular set $\set{C}$: we cannot even guarantee the \emph{almost global} convergence to the desired path by Corollary \ref{coroll_attra_P} and Corollary \ref{cor_nonattra_C}. However, by Theorem \ref{thm_dichotomy}, if the surface functions $\phi_i$ are \emph{known a priori}, then one can \emph{numerically} check whether the singular set $\set{C}$ is empty or not. If $\set{C} \ne \emptyset$, it obviously follows that global convergence to the desired path $\set{P}$ is not possible. But this requires the knowledge of the singular set, hence the knowledge of the specific expression of the vector field. 
	
	It is our interest to study whether we can obtain a general result about the convergence of the desired path \emph{without knowing the specific analytic expressions of the vector field and the singular set}. This interest is further motivated by the observation that the singular set is non-empty in many examples in the literature where the desired path is homeomorphic to the unit circle. Example \ref{simple_example1} is consistent with this observation. %

	\section{Singular points and non-path-converging trajectories} \label{sec_singular_pts}	
	\subsection{Existence of singular points} \label{doa_path}
	The following result shows that for the whole space to be the domain of attraction of an attractor, the shapes of the whole space and the attractor should be consistent topologically, independent of the system's specific form. 
	
	\begin{lemma} \label{lemma_deform_retract}
		Consider the autonomous system $\dot{x}=f(x)$, where $f$ is Lipschitz continuous and $x$ is defined on a smooth finite-dimensional manifold $\manifold$.  Suppose $\set{L}$ is a compact asymptotically stable embedded submanifold of $\manifold$. Then $\set{L}$ is a strong deformation retract of its domain of attraction.	
	\end{lemma}
	\begin{proof}
		Combine Theorem 5, Corollary 7 and Proposition 10 in \cite{moulay2010topological}.
	\end{proof}
	Rather than merely presenting this lemma, we list below several key steps behind this lemma, which are of theoretical interest. For convenience, the domain of attraction of $\set{L}$ is denoted by $\set{A}(\set{L})$.
	\begin{enumerate}[label=\textbf{S.\arabic*}, wide] %
		\item \label{enu1} 	As $\set{L}$ is a compact embedded submanifold, there exists a tubular neighborhood $\set{W}$ of $\set{L}$ which can be continuously shrunk to $\set{L}$; precisely, $\set{L}$ is a strong deformation retract of $\set{W}$.
		
		\item \label{enu2} 	The asymptotic stability of $\set{L}$ implies the existence of a smooth Lyapunov function $V$ on $\set{A}(\set{L})$ of which the time derivative $\dot{V}$ is negative definite on $\set{A}(\set{L})$ \cite[Theorem 3.1]{wilson1969smoothing}. We can choose a constant $\rho>0$ such that the sublevel set $V_\rho \defeq \{x \in \manifold : V(x) \le \rho \}$, which is a smooth manifold \cite{wilson1967structure}, is strictly contained in the tubular neighborhood\footnote{Existence of $\rho$ is guaranteed by the properties of the Lyapunov function $V$.} $\set{W}$. It follows from the negative definiteness of $\dot{V}$ that the vector field $f$ in Lemma \ref{lemma_deform_retract} only crosses the boundary $\partial  V_\rho$ of the sublevel set $V_\rho$ once, or precisely, the vector field is transverse to $\partial  V_\rho$. 
		
		\item \label{enu3}	Define the \emph{first hitting time} to be the first time instant when the system trajectory starting from $\bar{x} \in \set{A}(\set{L})$ reaches the sublevel set $V_\rho$. Then since the vector field is transverse to $\partial V_\rho$, it is proved that the \emph{first hitting time} $T_\rho(\bar{x})$ is continuous with respect to $\bar{x} \in \set{A}(\set{L})$. 
		
		\item \label{enu4}	The continuity of the first hitting time implies that $\set{A}(\set{L})$ can be continuously shrunk to $V_\rho$; precisely, $V_\rho$ is a strong deformation retract of $\set{A}(\set{L})$.
		
		\item \label{enu5}	The two continuous deformation processes in \ref{enu1} and \ref{enu4} imply that $\set{A}(\set{L})$ can be continuously shrunk to $\set{L}$. Or precisely, $\set{L}$ is a strong deformation retract of $\set{A}(\set{L})$.\footnote{Technically, \ref{enu1} implies that there exists a homotopy (corresponding to a strong deformation retraction) $s: \set{W} \times [0,1] \to \set{W}$ such that $s(w, 0)=w, s(w, 1) \in \set{L}$ for all $w \in \set{W}$ and $s(l, t) =l$ for all $l \in \set{L}$ and $t \in [0,1]$. Similarly, \ref{enu4} implies that there exists a homotopy (corresponding to a strong deformation retraction) $h: \set{A}(\set{L}) \times [0,1] \to \set{A}(\set{L})$ such that $h(y,0)=y, h(y,1) \in V_\rho$ for all $y \in \set{A}(\set{L})$ and $h(v, t)=v$ for all $v \in V_\rho$ and $t \in [0,1]$. A new homotopy (corresponding to a strong deformation retraction) $r: \set{A}(\set{L}) \times [0,1] \to \set{A}(\set{L})$ can be constructed as $r(y, t) = h(y, 2t)$ for $y \in \set{A}(\set{L})$ and $t \in [0, 1/2]$, and $r(y, t)=s(h(y,1), 2t-1)$ for $y \in \set{A}(\set{L})$ and $t \in (1/2, 1]$. This homotopy $r$ shows that $\set{L}$ is indeed a strong deformation retract of $\set{A}(\set{L})$ \cite[Theorem 4]{moulay2010topological}.}	
	\end{enumerate}

	\begin{remark}[Outline of an alternative proof] \label{remark_alternative_proof}
		We can independently derive the same result as Lemma \ref{lemma_deform_retract} using the ``local triviality'' property \cite[Chapter 10]{lee2015introduction} and the Wazewski set theorem \cite[Theorem 2.3]{conley1978isolated}. But due to the space limit, we only briefly introduce our proof technique for its theoretical interest. 
		
		The desired path $\set{P}$ (i.e., equivalent to $\set{L}$ in Lemma \ref{lemma_deform_retract}) being a compact regular level set implies the property called ``local triviality'' \cite[Chapter 10]{lee2015introduction}, which shows the ``stability'' of the topology of level sets near the desired path. This means that other level sets in the vicinity of the desired path look like the desired path, i.e. they are compact and homeomorphic to the desired path. Consequently, this vicinity is homeomorphic to the elliptic solid torus $\set{D} \times \mathbb{S}^1$, where $\set{D} \subseteq \mbr[n-1]$ is an ellipsoid centered at $\bm{0} \in \mbr[n-1]$ and is a sublevel set of the Lyapunov function \eqref{eqlyapunov}. Thus, we can indirectly study the properties of the original vector field $\vf$ in a neighborhood of the desired path by investigating the vector field $\vf'$ in this topological space: the elliptic solid torus $\set{D} \times \mathbb{S}^1$.  
		
		Then using the Lyapunov function \eqref{eqlyapunov} and its negative definite derivative, it can be shown that the vector field $\vf'$ is \emph{transverse} to the boundary $\set{S} \defeq \partial \set{D} \times \mathbb{S}^1$ of the elliptic solid torus $\set{D} \times \mathbb{S}^1$. This implies the continuity of the \emph{first hitting time} which is the first time instant a trajectory ``outside'' of the elliptic solid torus reaches the boundary $\set{S}$. Therefore, one may imagine that the set of all converging trajectories is continuously compressed into the boundary $\set{S}$, while every point in the boundary $\set{S}$ remains stationary during the whole continuous deformation process. This can be rigorously proved by the Wazewski set theorem \cite[Theorem 2.3]{conley1978isolated}, where the boundary $\set{S}$ turns out to be an \emph{exit set}, which roughly means that every trajectory starting from this boundary will \emph{immediately exit} from it. %
		
		Moreover, as the desired path is an embedded submanifold of $\manifold$, it has a tubular neighborhood $\set{W}$ and this neighborhood can be continuously shrunk onto the desired path \cite[Theorem 6.24, Proposition 6.25]{lee2015introduction}. Combining this continuous deformation process with the other one mentioned above, it is intuitive to see that the desired path $\set{P}$ is a (strong) deformation retract of its domain of attraction denoted by $\set{A}(\set{P})$.\footnote{Technically, due to the continuity of the first hitting time, we can define a continuous time function $T: \set{A}(\set{P}) \to \mathbb{R}$, such that $\Psi^{T(p)}(p)\in \set{W}$ for every $p \in \set{A}(\set{P})$, where $\Psi: \mathbb{R} \times \manifold \rightarrow \manifold$ is the flow of the dynamical system \eqref{eq1}. Then we can deform $\set{A}(\set{P})$ onto $\set{P}$ by first ``squeezing'' $\set{A}(\set{P})$ into $\set{W}$ via the homotopy $G(p, s)=\Psi^{2s \cdot T(p)}(p)$ with $s \in [0,\frac{1}{2}]$, and then $\set{W}$ onto $\set{P}$ via the homotopy $G(p, t)=H(\Psi^{T(p)}(p), 2t-1)$ with $t \in [\frac{1}{2},1]$, where $H$ is a homotopy corresponding to the deformation retraction of $\set{W}$ onto $\set{P}$. } 
		
		Our proof and that of Lemma \ref{lemma_deform_retract} both revolve around a) the regularity of the desired path (i.e., a compact regular level set and thus an embedded submanifold) and b) the continuity of the first hitting time (due to the transversality of the vector field to some surfaces encompassing the desired path). These two aspects turn out to be crucial in deriving the result.%
	\end{remark}
	\begin{remark}
		Note that Lemma \ref{lemma_deform_retract} does not hold if $\set{L}$ is not compact. In fact, the compactness of $\set{L}$ (or the desired path) is a crucial assumption in deriving the subsequent results. A counterexample is illustrated in \cite{yao2021doa}.
	\end{remark}

	An implication of Lemma \ref{lemma_deform_retract} is the following theorem: %
	\begin{theorem}[Homotopy equivalence] \label{thm_homotopy}
		The domain of attraction of  the desired path $\mathcal{P}$ with respect to \eqref{eq1} is homotopy equivalent to the unit circle $\mathbb{S}^1$.
	\end{theorem}
	\begin{proof}
		Let $\set{L}=\set{P}$ in Lemma \ref{lemma_deform_retract} and note that $\set{P}\approx \mathbb{S}^1$.
	\end{proof}

	We explain the potential utility of Theorem \ref{thm_homotopy}. One benefit relies on the computability of the related topological invariants. If global convergence to the limit cycle $\set{P}$ holds, then by Theorem \ref{thm_homotopy}, the configuration space $\manifold$ and the limit cycle $\set{P}$ are homotopy equivalent. This means that the homotopy equivalence of the configuration space $\manifold$ and the limit cycle $\set{P}$ is a necessary condition for the global convergence. This further implies that to check if global convergence to the limit cycle $\set{P}$ is possible, we can examine the topological invariants which are invariant under homotopy equivalences. These topological invariants include the Euler characteristic \cite[p. 178]{lee2010topologicalmanifolds}, homotopy groups \cite[p. 208]{lee2010topologicalmanifolds}, homology/cohomology groups \cite[pp. 339-355, pp. 374-378]{lee2010topologicalmanifolds}, etc. Some of these invariants are already known for some important topological spaces, and more are being investigated in the literature. For example, the fundamental group of $SO(3)$ is $\mathbb{Z}/2$, while that of $\mathbb{S}^1$ is $\mathbb{Z}$. \footnote{We thank W. Jongeneel for correcting a mistake in this expression made in an earlier version of the paper.} This implies that on the manifold $\manifold=SO(3)$, it is impossible to guarantee global convergence to a compact desired path $\set{P} \subseteq SO(3)$ which is homeomorphic to the unit circle. In practice, this implies that, for example, a quadcopter for which the orientations are defined on $SO(3)$, cannot follow a set of desired orientations defined by $\set{P} \approx \mathbb{S}^1$, from every initial orientation.
	
	Even though homotopy equivalent sets can look very different (compare $\mbr[2] \setminus \{\bm{0}\}$ and $\mathbb{S}^1$ for example), another benefit of Theorem \ref{thm_homotopy} is that it helps one obtain some intuition of how a domain of attraction looks. It also helps rule out some possibly wrong intuition that one might be misled into initially, especially when some path-converging or non-path-converging trajectories are of measure zero, and hence it is difficult, if not impossible, to be depicted by computer simulations. We will illustrate by examples in Section \ref{sec_simulation}. %

	The following theorem is particularized to $\manifold=\mbr[n]$. %
	\begin{theorem}[Impossibility of global convergence in $\mathbb{R}^{n}$] \label{thm_rn_global}
	If the configuration space $\manifold$ is the $n$-dimensional Euclidean space $\mbr[n]$, then global convergence to the desired path $\set{P}$ is not possible. In addition, if the dichotomy convergence property (i.e., Theorem \ref{thm_dichotomy}) holds globally\footnote{See Remark \ref{remark3} regarding when Theorem \ref{thm_dichotomy} globally holds.}, then the singular set $\set{C}$ is non-empty. 
	\end{theorem}
	\begin{proof}
	Since $\manifold=\mbr[n]$ is \emph{not} homotopy equivalent to\footnote{This can be seen from, for example, the fact that the Euler characteristic of $\mathbb{S}^1$ is $0$ while that of $\mbr[n]$ is $1$.} $\mathbb{S}^1$, the conclusion follows directly from Theorem \ref{thm_homotopy} and the dichotomy result of Theorem \ref{thm_dichotomy}. %
	\end{proof}
	\begin{remark}
		A different proof of the impossibility of global convergence of trajectories to desired paths homeomorphic to $\mathbb{S}^1$ is shown in our previous work \cite[Proposition 2]{yao2020singularity}, which is only applicable for guiding vector fields defined on the Euclidean space $\mbr[n]$.
	\end{remark}
	A major motivation of Theorem \ref{thm_rn_global} is the observation that in many examples in the literature \cite{kapitanyuk2017guiding,yao2018cdc,Goncalves2010,lawrence2008lyapunov}, the singular sets are non-empty, and hence global convergence to the desired path is simply not possible\footnote{When the singular set is non-empty, every trajectory of \eqref{eq1} starting from the singular set will simply remain stationary in the singular set (since a singular point is an equilibrium point of \eqref{eq1}), and thus it does \emph{not} converge to the desired path. Therefore, the \emph{global} convergence of trajectories to the desired path is simply \emph{not} possible.}. But is it the case that whenever $\set{P} \approx \mathbb{S}^1$, then there is always a non-empty singular set? For the 2D case, this is true by the Poincar{\' e}-Bendixson theorem, concluding that there always exist at least one singular point of the vector field within the region enclosed by the desired path, which is a limit cycle of the autonomous systems. Nevertheless, the Poincar{\' e}-Bendixson theorem cannot be straightforwardly extended to higher-dimensional spaces, for which the conclusion is not clear, but Theorem \ref{thm_rn_global} is able to give an affirmative answer. 
	
	It is true that if the singular set $\mathcal{C}$ is determined to be non-empty, then global convergence to the desired path $\mathcal{P}$ is not possible by Theorem \ref{thm_dichotomy}. However, the significance of Theorem \ref{thm_rn_global} is that it gives a more fundamental conclusion in the sense that it does not depend on the specific expressions of the surface functions $\phi_i$ (hence the vector field $\vf$), and avoids the possibly complicated computations of the singular set $\set{C}$ when the system dimensions are large. Most importantly, the independence of Theorem \ref{thm_rn_global} on the surface functions $\phi_i$ implies that even though we can choose different surface functions $\phi_i$ to represent the same desired path $\set{P}$, practically speaking, Theorem \ref{thm_rn_global} simply prevents us from hoping for better performance in terms of global convergence to the desired path by trying different surface functions $\phi_i$. %

	Note that the root causes of this topological obstacle are: a) the system \eqref{eq1} is autonomous; b) the stable desired path is homeomorphic to the unit circle while $\set{M}=\mbr[n]$. %
	Another implication of Theorem \ref{thm_rn_global} is that if one must achieve a global convergence result, then the only possible approach is to change the topology of the desired path $\set{P}$, if the autonomous system \eqref{eq1} is given. This is possible by, for example, ``tearing'' and ``stretching'' the compact desired path along an additional dimension, and thereby transforming it to an unbounded one which is homeomorphic to the real line (i.e., $\set{P} \approx \mbr[]$). This way, at least, the topological obstruction is removed. See \cite{yao2020mobile2,yao2020singularity} for our recent results along these lines applicable in Euclidean spaces.

	\subsection{The existence of non-path-converging trajectories}
	Theorem \ref{thm_rn_global} concludes that global convergence to the desired path in $\mbr[n]$ is not possible when $\set{M}=\mbr[n]$. Thus, it is of significant interest to show the existence of the non-path-converging trajectories, which do not converge to $\set{P}$. A theorem identifying some of these trajectories follows; the proof involves notions such as covering spaces, lifts, fundamental groups and homology, so we refer to Chapter 7, Chapter 11 and Chapter 13 in \cite{lee2010topologicalmanifolds} for an introduction.%

	Before presenting the main theorem, we first provide some preliminary lemmas and notations. The notation $H_{n-1}(\cdot)$ denotes the $(n-1)$-dimensional homology group \cite[Chapter 13]{lee2010topologicalmanifolds}, and $(\cdot)_{*}$ denotes the homomorphism between homology groups induced by a continuous map $(\cdot)$.  Let $\set{B}^{n-1} \defeq \{x \in\mbr[n-1] : \norm{x} < 1\}$ be the unit open ball in $\mbr[n-1]$ centered at $\bm{0}$, and $\set{B}_{-}^{n-1} \defeq \set{B}^{n-1}\setminus\{\bm{0}\}$.
	\begin{lemma} \label{lemma_trivialhomology}
		There holds $H_{n-1}(\set{B}_{-}^{n-1}\times\mathbb{R})=\{0\}.$
	\end{lemma}
	\begin{proof}
		Since $\mathbb{S}^{n-2}$ is a deformation retract of $\set{B}_{-}^{n-1}$, and $\mathbb{R}$ is contractible (i.e., homotopy equivalent to a singleton),  $\set{B}_{-}^{n-1}\times\mathbb{R}$ is homotopy equivalent to $\mathbb{S}^{n-2} \times \{x_{0}\}$, for a point $x_{0} \in \mathbb{R}$. In addition, $\mathbb{S}^{n-2} \times \{x_{0}\}$ is homeomorphic to $\mathbb{S}^{n-2}$. Therefore, we have 
		$
		H_{n-1}(\set{B}_{-}^{n-1}\times\mathbb{R})=H_{n-1}(\mathbb{S}^{n-2})=\{0\}
		$ \cite[Theorem 13.23]{lee2010topologicalmanifolds}.
	\end{proof}

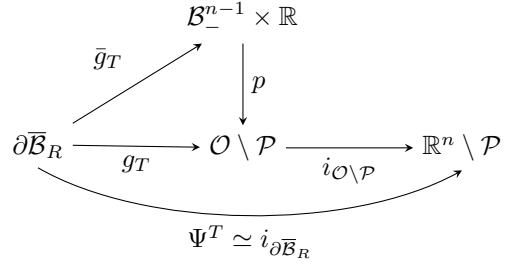
\begin{figure}[tb]
		\centering
		\begin{tikzpicture}
		\matrix (m) [matrix of math nodes,row sep=3em,column sep=4em,minimum width=2em]
		{
			{} & \set{B}_{-}^{n-1} \times \mbr[] & {}\\
			\partial \overline{\set{B}}_R & \set{O}\setminus \set{P} & \mbr[n] \setminus \set{P} \\};
		\path[-stealth]
		(m-2-1) edge node [below] {$g_{T}$} (m-2-2) 
		(m-1-2) edge node [right] {$p$} (m-2-2) 
		(m-2-1) edge node [above left=2] {$\bar{g}_{T}$} (m-1-2)
		(m-2-2) edge node [below] {$i_{\set{O}\setminus \set{P}}$} (m-2-3);
		\draw [->, >=stealth, out=-30, in=-150, looseness=0.75] (m-2-1.south) to node[below]{$\Psi^T \simeq i_{\partial \overline{\set{B}}_R}$} (m-2-3.south);
		\end{tikzpicture}
		\caption{Relations of continuous maps in Theorem \ref{thm_locating_diverging}. The set $\partial \overline{\set{B}}_R$ is an $(n-1)$-dimensional sphere. The map $p$ is a covering map, $\bar{g}_{T}$ is a lift of $g_{T}$ such that $g_{T}=p \circ \bar{g}_{T}$, and $\Psi^T = i_{\set{O} \setminus \set{P}} \circ g_{T}$, where $i_{\set{O} \setminus \set{P}}$ is an inclusion map. The map $\Psi^T$ is homotopic to the inclusion map $i_{\partial \overline{\set{B}}_R}$. }
		\label{fig: thm4}
	\end{figure}

	\begin{lemma} \label{lemma_iznontrivial}
		Let $i_{\partial \overline{\set{B}}_R}:\partial \overline{\set{B}}_R\rightarrow\mathbb{R}^{n} \setminus \mathcal{P}$ be the inclusion map, where $\partial \overline{\set{B}}_R$ denotes the boundary of a closed ball $\overline{\set{B}}_R \subseteq \mbr[n]$ of radius $R$ containing the desired path $\set{P}$. The homomorphism $i_{\partial \overline{\set{B}}_R,*}: H_{n-1}(\partial \overline{\set{B}}_R) \to H_{n-1}(\mbr[n] \setminus \set{P})$ induced by $i_{\partial \overline{\set{B}}_R}$  is non-trivial\footnote{A homomorphism $h:G \to G'$ between any two groups (e.g., fundamental groups, homology groups) is called \emph{trivial} if $h$ maps every element in $G$ to the identity element, denoted by $0$, in $G'$. The homomorphism $h$ is called \emph{non-trivial} if it is not trivial. }.
	\end{lemma}
	\begin{proof}
		Fix $x_{0}\in \set{P}$, and denote by $j_{\partial \overline{\set{B}}_R}$ and $j_{\mathbb{R}^{n}\setminus\set{P}}$ respectively the inclusions of $\partial \overline{\set{B}}_R$ and $\mathbb{R}^{n}\setminus\set{P}$ into $\mathbb{R}^{n}\setminus\{x_{0}\}$. Then 
		\begin{equation} \label{eq_inclusion}
		j_{\partial \overline{\set{B}}_R}=j_{\mathbb{R}^{n}\setminus\set{P}}\circ i_{\partial \overline{\set{B}}_R}.
		\end{equation}
		Since $x_{0}$ lies inside the ball $\overline{\set{B}}_{R}$ of which the boundary is $\partial \overline{\set{B}}_R$, $\partial \overline{\set{B}}_R$ is a deformation retract of $\mathbb{R}^{n}\setminus\{x_{0}\}$ and hence $j_{\partial \overline{\set{B}}_R}$ is a homotopy equivalence between $\partial \overline{\set{B}}_R$ and $\mathbb{R}^{n}\setminus\{x_{0}\}$. Thus $j_{\partial \overline{\set{B}}_R}$ induces an isomorphism $j_{\partial \overline{\set{B}}_R,*}$ between $H_{n-1}(\partial \overline{\set{B}}_R)$ and $H_{n-1}(\mathbb{R}^{n}\setminus\{x_{0}\})$ \cite[Corollary 13.9]{lee2010topologicalmanifolds}. Since \eqref{eq_inclusion} implies \cite[Proposition 13.2]{lee2010topologicalmanifolds}
		\begin{align*}
		&j_{\partial \overline{\set{B}}_R,*} : H_{n-1}(\partial \overline{\set{B}}_R) \to H_{n-1}(\mathbb{R}^{n}\setminus\{x_{0}\}) \\
		&j_{\partial \overline{\set{B}}_R,*}=j_{\mathbb{R}^{n}\setminus\set{P},*}\circ i_{\partial \overline{\set{B}}_R,*},
		\end{align*}
		and $H_{n-1}(\partial \overline{\set{B}}_R) \simeq H_{n-1}(\mathbb{R}^{n}\setminus\{x_{0}\}) \simeq\mathbb{Z}$ \cite[Theorem 13.23]{lee2010topologicalmanifolds}, this implies that $i_{\partial \overline{\set{B}}_R,*}$ is non-trivial. 
	\end{proof}
	Now we are ready to state the following main theorem.
	\begin{theorem}[Existence of non-path-converging trajectories] \label{thm_locating_diverging}
		Suppose $n \ge 3$ for the autonomous differential equation \eqref{eq1}, where the desired path $\set{P} \subseteq \mbr[n]$ is an embedded submanifold in $\mbr[n]$ and a (locally) asymptotically stable limit cycle. For any closed ball $\overline{\set{B}}_R \subseteq \mbr[n]$ that contains $\set{P}$ (precisely, $\set{P} \subseteq \interior \overline{\set{B}}_R$), there exists at least one trajectory of \eqref{eq1} starting from the boundary $\partial \overline{\set{B}}_R$ of the ball $\overline{\set{B}}_R$ that does not converge to $\set{P}$.
	\end{theorem}
	\begin{proof}
		\textbf{Step 1 (construct the map $\Psi^T$):} We prove by contradiction. Denote by $\Psi : \mbr[] \times \mbr[n] \to \mbr[n]$ the flow of \eqref{eq1}; i.e. $\gamma(t)=\Psi^{t}(x)$ is the solution of \eqref{eq1} with the initial condition $\gamma(0)=x$.   Suppose that every trajectory starting from the boundary $\partial \overline{\set{B}}_R$ of the closed ball $\overline{B}_R$ (i.e., $\partial \overline{\set{B}}_R$ is an $(n-1)$-dimensional sphere) converges to the limit cycle $\set{P}$; then for any point $x \in \partial \overline{\set{B}}_R$, there exists some $T_{x}>0$ such that $\Psi^{t}(x)\in \set{O}$ for all $t>T_{x}$, where $\set{O}$ is a tubular neighborhood $\set{O}$ of $\set{P}$ in $\mbr[n]$. \footnote{Since $\set{P}$ is an embedded submanifold in $\mbr[n]$, a tubular neighborhood $\set{O}$ of $\set{P}$ always exists \cite[Theorem 6.24]{lee2015introduction}.} By the compactness of $\partial \overline{\set{B}}_R$ and the asymptotic stability of the limit cycle  $\set{P}$, one can show that there exists $T>0$ such that $\Psi^{T}(\partial \overline{\set{B}}_R)\subseteq \set{O} \setminus \set{P}$. 
		
		\textbf{Step 2 (rewrite $\Psi^T$):} We can write $\Psi^T: \partial \overline{\set{B}}_R \to \mbr[n] \setminus \set{P}$ as the composition of two functions as below:
		\[
			\Psi^T=i_{\set{O} \setminus \set{P}} \circ g_{T},
		\]
		where $g_{T}:\partial \overline{\set{B}}_R\rightarrow\mathcal{O} \setminus \mathcal{P}$ is simply the codomain restriction of $\Psi^T$, and $i_{\set{O} \setminus \set{P}}: \mathcal{O} \setminus \mathcal{P}\rightarrow\mathbb{R}^{n} \setminus \mathcal{P}$ is the inclusion map. 
	
		\textbf{Step 3 (construct a covering map):} Since $\set{P}$ is an embedded submanifold in $\mbr[n]$, and $\set{O}$ is a tubular neighborhood, there exists a diffeomorphism\footnote{More precisely, the tubular neighborhood $\set{O}$ being diffeomorphic to $\set{B}^{n-1}\times \mathbb{S}^1$ is because the normal bundle of a loop in $\mathbb{R}^n$ is orientable.} $\beta$ from $\set{O}$ to $\set{B}^{n-1}\times \mathbb{S}^1$ such that $\beta(\set{P})=\{\bm{0}\} \times \mathbb{S}^1$  \cite[Theorem 6.24]{lee2015introduction}. Since $\beta$ is a diffeomorphism between $\set{O}$ and $\set{B}^{n-1}\times \mathbb{S}^{1}$, and $\beta(\set{P})=\{\bm{0}\}\times \mathbb{S}^{1}$, it follows that $\set{O}\setminus\set{P}$ is diffeomorphic to $\set{B}_{-}^{n-1}\times \mathbb{S}^{1}$ with a diffeomorphism $\beta': \set{O} \setminus \set{P} \to \set{B}_{-}^{n-1} \times \mathbb{S}^{1}$, where $\beta'(x)=\beta(x)$ for $x \in \set{O}\setminus\set{P}$. Therefore, 
		\begin{align*}
		&p: \set{B}_{-}^{n-1}\times\mathbb{R}\rightarrow \set{O} \setminus \set{P} \\
		&(u,\theta) \mapsto \beta'^{-1} \big( (u,e^{i\theta}) \big)
		\end{align*}
		is a covering map \cite[p. 278]{lee2010topologicalmanifolds} with $\set{B}_{-}^{n-1}\times\mathbb{R}$ being the covering space of $\set{O} \setminus \set{P}$. 
		
		\textbf{Step 4 (lift $g_T$ and form a contradiction):} Since $\partial \overline{\set{B}}_R$ is simply connected and locally path-connected for $n \ge 3$ \cite[Theorem 7.20]{lee2010topologicalmanifolds}, the continuous map $g_{T}: \partial \overline{\set{B}}_R \to \set{O} \setminus \set{P}$ can be lifted \cite[Corollary 11.19]{lee2010topologicalmanifolds}; that is, there exists $\bar{g}_{T}: \partial \overline{\set{B}}_R \to \set{B}_{-}^{n-1} \times \mbr[] $ such that $g_{T}=p \circ \bar{g}_{T}$ (see Fig. \ref{fig: thm4}). It follows from Lemma \ref{lemma_trivialhomology} that $\bar{g}_{T,*}:  H_{n-1}(\partial \overline{\set{B}}_R)\rightarrow H_{n-1}(\set{B}_{-}^{n-1} \times \mathbb{R})$ is trivial. Therefore, $\Psi^T_*=i_{\set{O} \setminus \set{P}, *} \circ p_{*}\circ\bar{g}_{T,*}$ is trivial \cite[Proposition 13.2]{lee2010topologicalmanifolds}. Let $i_{\partial \overline{\set{B}}_R}: \partial \overline{\set{B}}_R \to \mathbb{R}^n \setminus \set{P}$ be the inclusion map of $\partial \overline{\set{B}}_R$ to $\mathbb{R}^n \setminus \set{P}$. Since $i_{\partial \overline{\set{B}}_R}:\partial \overline{\set{B}}_R\rightarrow\mathbb{R}^{n} \setminus \mathcal{P}$ and $\Psi^{T}: \partial \overline{\set{B}}_R\rightarrow\mathbb{R}^{n} \setminus \mathcal{P}$ are homotopic\footnote{A homotopy between $i_{\partial \overline{\set{B}}_R}:\partial \overline{\set{B}}_R\rightarrow\mathbb{R}^{n} \setminus \mathcal{P}$ and $\Psi^{T}: \partial \overline{\set{B}}_R\rightarrow\mathbb{R}^{n} \setminus \mathcal{P}$ is	$G: \partial \overline{\set{B}}_R\times[0,1]\rightarrow\mathbb{R}^{n} \setminus \mathcal{P}$ defined by $G(x,s)=\Psi^{s\cdot T}(x)$. }, it follows that $i_{\partial \overline{\set{B}}_R, *}=\Psi^T_{*}$ \cite[Theorem 13.8]{lee2010topologicalmanifolds}, and hence $i_{\partial \overline{\set{B}}_R, *}$ is trivial. However, the conclusion that $i_{\partial \overline{\set{B}}_R,*}$ is trivial contradicts Lemma \ref{lemma_iznontrivial}.	
	\end{proof}	
	\begin{remark}[Why require $n \ge 3$?]
		The reason that the theorem cannot be proved when $n=2$ (at least following the argument for $n \geq 3$) is that the continuous map $g_T$ cannot be lifted (i.e., there does not exist the continuous map $\bar{g}_T$ such that $g_T = p \circ \bar{g}_T$). Note that $\partial \overline{\set{B}}_R \approx \mathbb{S}^{n-1}$. If $n=2$, then $\partial \overline{\set{B}}_R \approx \mathbb{S}^{1}$, which is \emph{not} simply connected. Then \cite[Corollary 11.19]{lee2010topologicalmanifolds} cannot be used to imply the existence of the lift $\bar{g}_T$ of the continuous map $g_T$ as what we did in the proof. In fact, it can be further shown that this lift $\bar{g}_T$ does not exist according to the lifting criterion \cite[Theorem 11.18]{lee2010topologicalmanifolds}. This is elaborated as follows. First, the fundamental group of $\partial \overline{\set{B}}_R$ is $\pi_1(\partial \overline{\set{B}}_R, \rho_1)=\pi_1(\mathbb{S}^1, \rho_2)=\mathbb{Z}$ for any base points $\rho_1 \in \partial \overline{\set{B}}_R$ and $\rho_2 \in \mathbb{S}^1$. Second, when $n=2$, $\set{B}_{-}^{n-1} \times \mbr[]=( (-1,0) \cup (0,1) ) \times \mbr[]= \big( (-1,0) \times \mbr[] \big) \cup \big( (0,1) \times \mbr[] \big) \defeq \set{A}_1 \cup \set{A}_2$, which consists of two disjoint contractible subspaces denoted by $\set{A}_1$ and $\set{A}_2$. Therefore, the fundamental group of $\set{A}_1$ and $\set{A}_2$ at any of their base points is $0$ in both cases. These two facts imply that the conditions of \cite[Theorem 11.18]{lee2010topologicalmanifolds} cannot be satisfied (since the homomorphism $g_{T,*}$ between corresponding fundamental groups is non-trivial while the homomorphism $p_{*}$ is trivial). Therefore, there does not exist the lift $\bar{g}_T$ of $g_T$, and the subsequent proof cannot proceed. In fact, one can easily observe using an example that the theorem indeed does not hold for the case where $n=2$ (e.g., see the first example in Section \ref{sec_simulation}).
	\end{remark}	
	\begin{remark}
	The sphere $\partial \overline{\set{B}}_R$ in the theorem can be generalized to any smooth $(n-1)$-sphere (i.e., any smooth submanifold diffeomorphic to $\mathbb{S}^{n-1}$). This is due to the Jordan-Brouwer Separation theorem \cite[Chapter 2.5]{guillemin2010differential} and the Generalized Schoenflies Theorem \cite[Chapter V]{bing1983geometric}, \cite{brown1960proof,mazur1959embeddings}.
	\end{remark}

	The essential idea behind the proof of Theorem \ref{thm_locating_diverging} is that the ball $\overline{\set{B}}_R$ cannot be continuously shrunk to a point if there are ``holes'' or ``obstacles'' in the ball's interior. Imagine a two-dimensional sphere containing the desired path, which can be treated as a circle, and the whole space is $\mbr[3]$. The sphere will be continuously shrunk as its points move towards the desired path along the system's flows. Note that the deforming sphere does not intersect with the desired path during this process since the path is a periodic orbit of the autonomous system (i.e., the desired path can be seen as an obstacle). If \emph{all} points of the sphere are driven by the flows to converge to the path ultimately, the deforming sphere will be pulled apart into pieces, but this is impossible due to the continuity of flows. This essential idea might be utilized to show the following two conjectures. 
	
	\begin{conj} \label{conjecture1}
		Suppose the assumptions of Theorem \ref{thm_locating_diverging} hold, but the domain of the vector field becomes $\mbr[n] \setminus \set{H}$, where $\set{H}$ is a non-empty set. If there exists a closed ball $\overline{\set{B}}_R$ such that $\set{H} \cap \interior \overline{\set{B}}_R \ne \emptyset$, then there exists at least one trajectory of \eqref{eq1} starting from the boundary of the ball $\partial \overline{\set{B}}_R$ such that it does not converge to the limit cycle $\set{P}$.		
	\end{conj}
	Here the set $\set{H}$ can be regarded as a collection of holes, and the condition $\set{H} \cap \interior \overline{\set{B}}_R \ne \emptyset$ means that there is at least one hole inside the closed ball $\overline{\set{B}}_R$. Furthermore, as this ball can be shrunk to be arbitrarily close to the ``hole'', this indicates that in the close vicinity of the ``hole'', such a non-path-converging trajectory exists. This somehow gives a way to locate where the non-path-converging trajectories originate from. In practice, this indicates the following conjecture:
	\begin{conj} \label{conjecture2}
		It is impossible for vehicles (e.g. a wheeled robot, a drone) of which the motions are governed by the autonomous system \eqref{eq1} to smoothly converge to a desired configuration (e.g., position, orientation) from every initial configuration in an environment scattered with obstacles.
	\end{conj}
	This is because the vehicles cannot bump into obstacles, and thus these obstacles are regarded as ``holes'' $\set{H}$ in the configuration space $\set{M}$ (e.g., $\set{M}=\mbr[2]$ for a wheeled robot moving on a plane or $\set{M}=SE(3)$ for a drone flying with different poses). This is a general conjecture independent of how the configuration space looks; a similar conclusion was drawn in \cite{RimonKoditschek1992,koditschek1990robot} for the special case of a \textit{sphere world}. Conjecture \ref{conjecture2} is also consistent with the recent result in \cite[Theorem 11]{Braun2017OnE}, which shows that the global asymptotic stabilization of the origin using continuous controllers is not possible in environments with (bounded) obstacles.

	\section{Numerical simulations} \label{sec_simulation}
	
	We present simulation examples in this section to illustrate the theoretical results in Section \ref{doa_path}. %
	These examples are meant to display the initial conditions under which trajectories do \emph{not} converge to the desired path. They are consistent with the claim that the domains of attraction of the desired path are indeed homotopy equivalent to the unit circle. %
	
\begin{figure}[tb]
		\centering
		\includegraphics[width=0.6\linewidth]{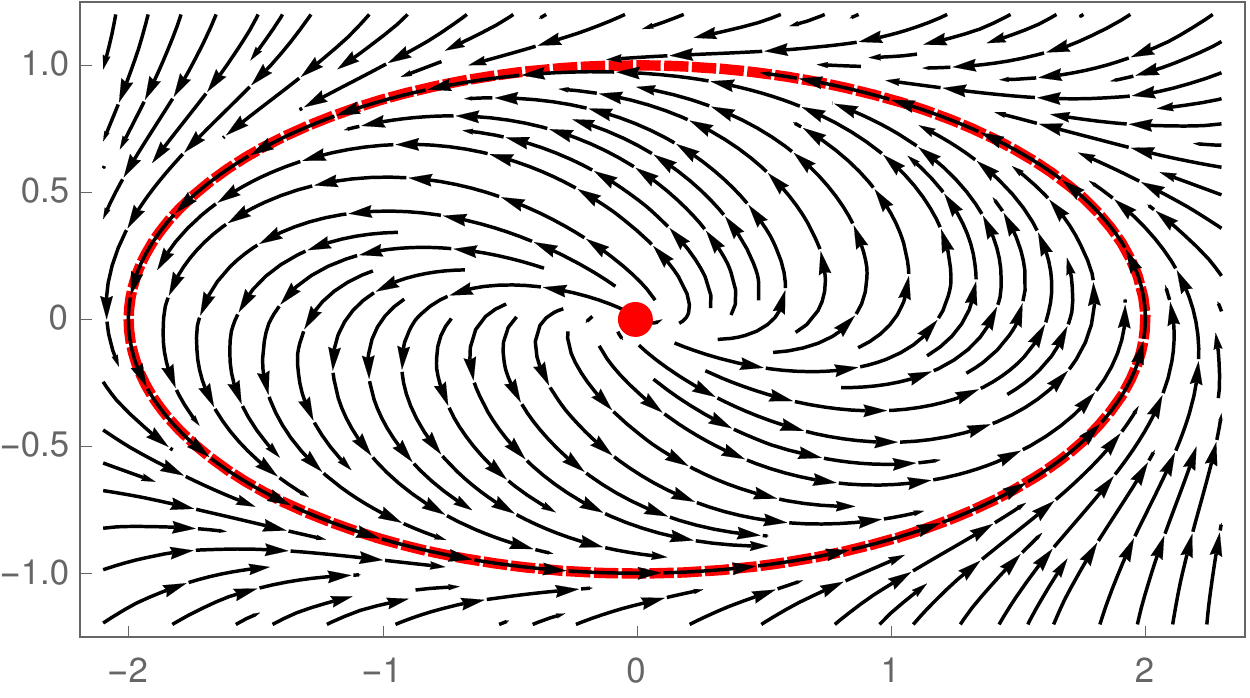}
		\caption{Streamlines of the first example. The solid line (red) is the desired path, and the center point (red) is the only singular point. }
		\label{fig:ellipsestreamline}
	\end{figure}

	In the first example, we choose $\phi=x^2 / 4+y^2-1$ and the gain is $k=1$ in \eqref{gvfe}. For the 2D case, the wedge product in \eqref{gvfe} is calculated by $\orthoterm = E \phi$, where $E \in SO(2)$ is a $90^\circ$ rotation matrix. It can be numerically calculated that there is only one singular point $s_{A1}=(0,0)$. Since the eigenvalues of the Jacobian matrix of the vector field at this singular point have all positive real parts, the singular point is a source of \eqref{eq1}. Therefore, one could conclude that the domain of attraction of the ellipse is $\mathcal{A}_1 \defeq \mbr[2] \setminus s_{A1}$, and it is indeed homotopy equivalent to the unit circle (see the streamlines in Fig. \ref{fig:ellipsestreamline}). 
	
	\begin{figure}[tb]
	\centering
	\subfigure[]{
		\includegraphics[width=0.5\linewidth]{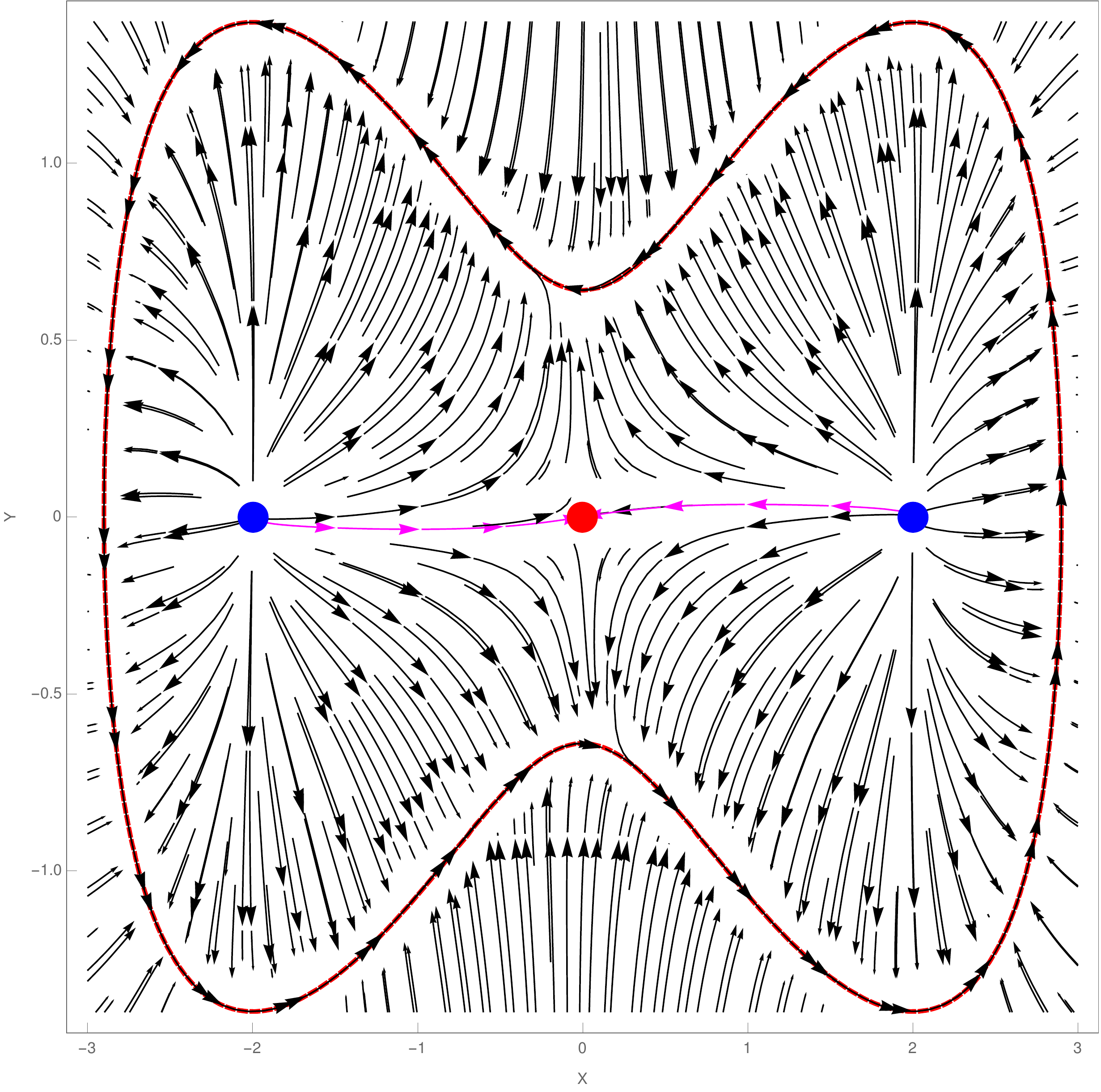}
		\label{fig:cassiniovalstreamline}
	}%
	\subfigure[]{
		\includegraphics[width=0.5\linewidth]{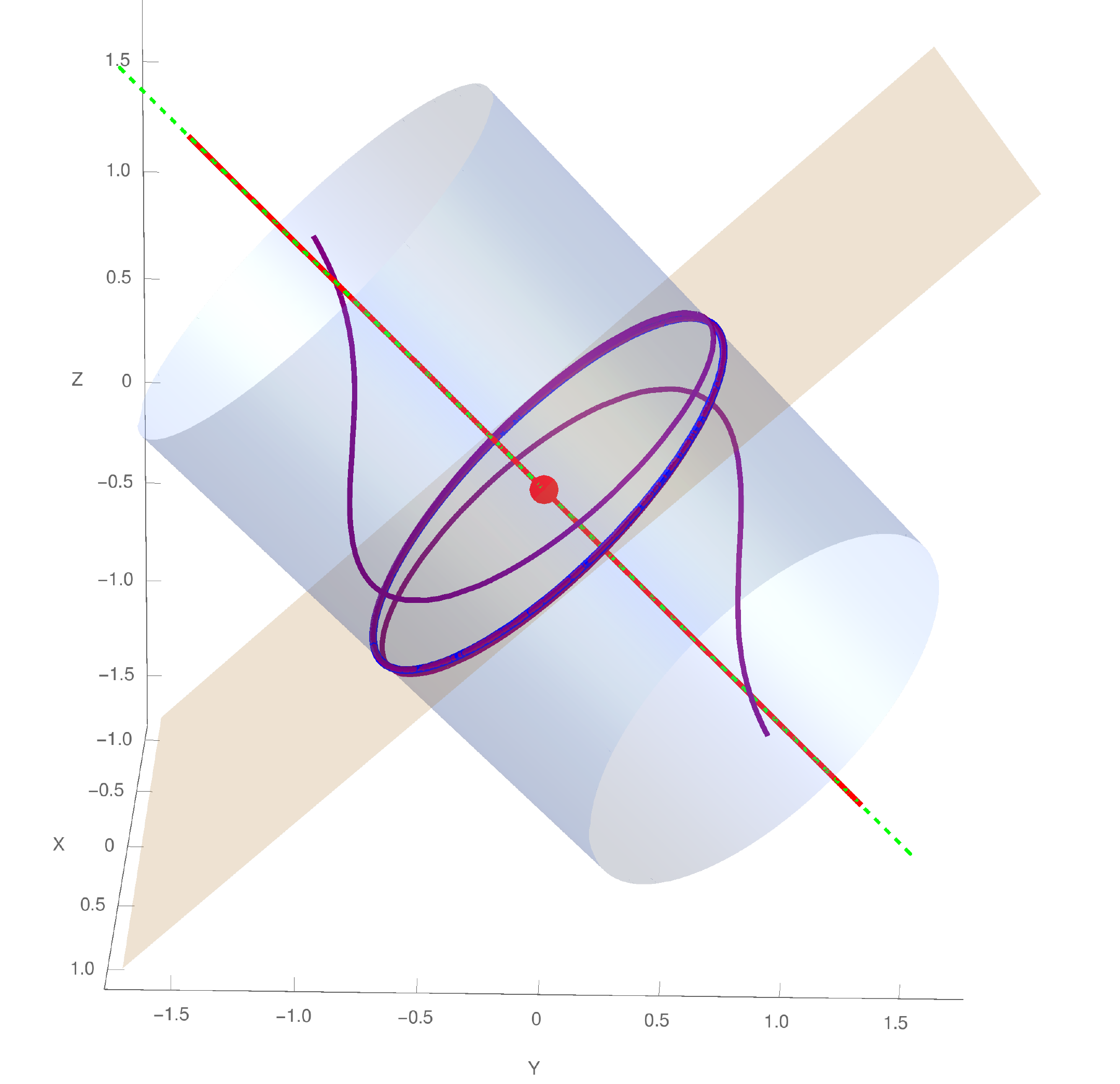}
		\label{fig:3dcircletraj}
	}
	\caption{\subref{fig:cassiniovalstreamline} Streamlines of the second example. The solid line (red) is the desired path and the three point (two blue and one red) are singular points. The magenta streamlines converge to the central singular point. \subref{fig:3dcircletraj} The 3D tilted circle (blue dashed line) generated by the intersection of a rotated (i.e., non axis-parallel) right circular cylinder and a rotated plane. The point (red) at the center is the only singular point. The dashed line (green) is the cylinder's line of symmetry and is normal to the plane. The four solid lines (two red and two purple) are trajectories of \eqref{eq1} with different initial conditions.}
	\label{fig:exmples}
	\end{figure}

	In the second example, we choose a Cassini oval, which is characterized by $\phi=x ^4 + y ^4 - 2 a^2(x^2 - y ^2) + a^4 - b^4$, where $a=2,b=2.1$. The gain of the vector field \eqref{gvfe} is $k=0.1$. One could calculate that there are three singular points: $s_{B1}=(0,0), s_{B2}=(a,0),s_{B3}=(-a,0)$. Following the first example, one might think that the domain of attraction of this example is $\mathcal{A}_2 \defeq \mbr[2] \setminus \{s_{B1},s_{B2},s_{B3}\}$, similar to the case of a circle or an ellipse. A direct computer numerical calculation may also lead one to this conclusion\footnote{For example, by invoking \texttt{StreamPlot} in \texttt{Mathematica} directly.}. But this is incorrect by Theorem \ref{thm_homotopy}, as $\mathcal{A}_2$ is \emph{not} homotopy equivalent to $\mathbb{S}^1$. In fact, by examining the eigenvalues of the Jacobian matrices at these three singular points, we find out that $s_{B1}$ is a saddle point while $s_{B2}$ and $s_{B3}$ are unstable sources. By the Hartman-Grobman theorem \cite[Theorem 7.3]{sastry2013nonlinear}, one could conclude that there must be a ``line'' of points (the stable manifold of $s_{B1}$) starting from which the trajectories of \eqref{eq1} converge to the saddle point $s_{B1}$, and by Theorem \ref{thm_homotopy}, this ``line'' probably connects the other two singular points. In fact, after experimenting with different plotting parameters\footnote{We use \texttt{StreamPlot} in \texttt{Mathematica} and combine figures with different sampling parameters.}, we find out that there indeed exists such a ``line'' $\mathcal{L}_{B}$ (see the magenta arrows in Fig. \ref{fig:cassiniovalstreamline}). Then the domain of attraction of the Cassini oval is $\set{A}_2 \defeq \mbr[2] \setminus \mathcal{L}_{B} \approx \mathbb{S}^1$.

	In the third example, a 3D tilted circle is the desired path, which is the intersection of a rotated (i.e., non axis-parallel) right circular cylinder and a rotated plane described by $\phi_1 = x^2 + 0.5(y + z)^2 - 1=0$ and $\phi_2 = y - z=0$ respectively (see Fig. \ref{fig:3dcircletraj}). The gains are chosen as $k_1=k_2=1$ for the vector field \eqref{gvfe}. There is only one singular point at the origin $s_{C1}=(0,0,0)$, and there is only one eigenvalue of the Jacobian matrix with a positive real part. The line of symmetry of the cylinder, denoted by $\mathcal{L}_{C} \defeq \{(0,u,-u) \in \mbr[3]: u \in \mbr[]\}$, is normal to the plane. The vector field evaluated at any point $p \in \mathcal{L}_{C}$ on this line is $\vf(p)=(0, -2y, 2y)$, where $y$ is the second coordinate of the point. This vector $\vf(p)$ aligns with $\mathcal{L}_{C}$ and points towards the singular point $s_{C1}$. This is consistent with Theorem \ref{thm_locating_diverging}, since for every ball that contains that desired path, every trajectory starting from the intersection point of the ball  and the line of symmetry $\set{L}_C$ will move along $\set{L}_C$ and converge to the singular point rather than the desired path. In addition, the domain of attraction of the tilted circle is $\mathcal{A}_3 \defeq \mbr[3] \setminus \mathcal{L}_{C}$, which is also homotopy equivalent to $\mathbb{S}^1$. This example is the same as that in \cite{lawrence2008lyapunov}. Note that, \cite{lawrence2008lyapunov} cannot claim global convergence, since the initial condition is restricted in a compact set. As shown here, only \emph{almost global} convergence to the 3D circle can be achieved. %
	
\begin{figure}[tb]
	\centering
	\newcommand{\nvar}[2]{%
		\newlength{#1}
		\setlength{#1}{#2}
	}
	
	\nvar{\dg}{0.25cm}
	\def\dw{0.3}\def\dh{0.2}\def\dl{0.3}
	\def\link{\draw[double distance=0.8mm, very thick] (0,0)--}
	\def\joint{%
		\filldraw [fill=white] (0,0) circle (4pt);
		\fill[black] circle (2pt);
	}
	\def\grip{%
		\draw[thick](0cm,\dg)--(0cm,-\dg);
		\draw[ultra thick] (0cm, 0.5\dg) -- +(0.6\dg,0cm);
		\draw[ultra thick] (0cm, -0.5\dg) -- +(0.6\dg,0cm);
		\fill[black] circle (2pt);
	}
	
	\def\robotbase{%
		\draw[rounded corners=6pt] (-\dw,-\dh)--(-\dw, 0)--(0,\dh)--(\dw,0)--(\dw,-\dh);
		\draw (-0.5,-\dh)--(0.5,-\dh);
		\fill[pattern=north east lines] (-0.5,-\dl) rectangle (0.5,-\dh);
	}
	
	\newcommand{\xyframe}[2]{%
		\draw[-latex, line width=1pt] (0,0) -- (#1,0) node[below=2]{$x$};
		\draw[-latex, line width=1pt] (0,0) -- (0,#2) node[left]{$y$};
	}
	
	\nvar{\ddx}{1cm}
	\newcommand{\angann}[2]{%
		\begin{scope}[red]
			\draw [dashed, red] (0,0) -- (1.2\ddx, 0pt);
			\draw [-latex, shorten >=3.5pt] (\ddx,0pt) arc (0:#1:\ddx);
			\node at (#1/2-2:\ddx+8pt) {#2};
		\end{scope}
	}
	
	\newcommand{\lineann}[4][0.5]{%
		\begin{scope}[rotate=#2, blue,inner sep=2pt]
			\draw[dashed, blue!40] (0,0) -- +(0,#1)
			node [coordinate, near end] (a) {};
			\draw[dashed, blue!40] (#3,0) -- +(0,#1)
			node [coordinate, near end] (b) {};
			\draw[|<->|] (a) -- node[fill=white] {#4} (b);
		\end{scope}
	}
	
	\newcommand{\twolink}[4]{%
		\robotbase
		\link(#1:#2);
		\joint
		\begin{scope}[shift=(#1:#2), rotate=#1]
			\link(#3:#4);
			\joint
			\begin{scope}[shift=(#3:#4), rotate=#3]
				\grip
			\end{scope}
		\end{scope}
	}	
	\def\thetaone{60}
	\def\Lone{2}
	\def\thetatwo{30}
	\def\Ltwo{3}
	
	\subfigure[]{
		\begin{tikzpicture}[scale=0.7]
			\xyframe{1.8}{1.8}
			\robotbase
			\angann{\thetaone}{$\theta_1$}
			\lineann[0.6]{\thetaone}{\Lone}{$L_1$}
			\link(\thetaone:\Lone);
			\joint
			\begin{scope}[shift=(\thetaone:\Lone), rotate=\thetaone]
			\angann{\thetatwo}{$\theta_2$}
			\lineann[-1.6]{\thetatwo}{\Ltwo}{$L_2$}
			\link(\thetatwo:\Ltwo);
			\joint
			\begin{scope}[shift=(\thetatwo:\Ltwo), rotate=\thetatwo]
			\grip
			\end{scope}
			\end{scope}
			
			\begin{scope}[dash pattern=on 4pt off 1pt]
			\twolink{200}{\Lone}{-110}{\Ltwo}
			\end{scope}
			
			\draw[thick, dashed] (0,\Ltwo) circle (\Lone);
			\fill[black] (0,\Ltwo) circle (2pt);
		\end{tikzpicture}
		\label{fig: robot_arm}
	}
	\subfigure[]{
		\includegraphics[width=0.5\columnwidth]{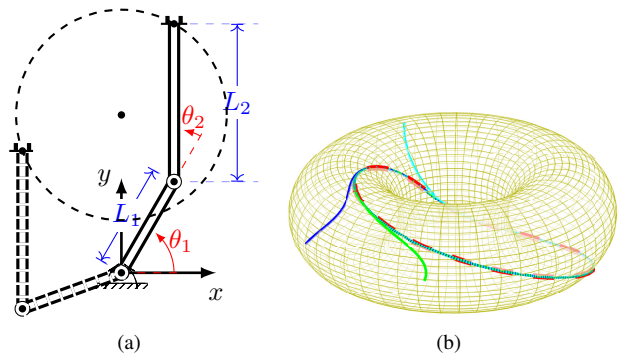} %
		\label{fig:torusrobotarm}
	}
	\caption{\subref{fig: robot_arm} A planar robot arm with two revolute joints of which the angles are denoted by $\theta_1$ and $\theta_2$. The dashed line is the desired path to follow in the Cartesian space $\mbr[2]$, which corresponds to $\phi(\theta_1,\theta_2)=\theta_1+\theta_2-\pi/2=0$ in the joint space $\manifold=\mathbb{S}^1 \times \mathbb{S}^1$. \subref{fig:torusrobotarm} Simulations of trajectories on the torus $\manifold=\mathbb{S}^1 \times \mathbb{S}^1$. The torus $\manifold=\mathbb{S}^1 \times \mathbb{S}^1$ is transparent such that trajectories are visible. The red dashed line is the desired path $\set{P} \subseteq \manifold$. The blue, green and cyan solid lines are trajectories corresponding to initial conditions $(\theta_1, \theta_2)=(0,0), (0.3\, \pi,0)$ and $(1.5\,\pi,0.5\,\pi)$ respectively.}
	\end{figure}

	In the fourth example, we consider a configuration space which is \emph{not} the Euclidean space. Specifically, we consider the example of a planar robot arm with two revolute joints, and we want to control the angles $\theta_1, \theta_2 \in \mathbb{S}^1$ of these two revolute joints such that the end-effector of the robot arm follows some desired trajectory (see Fig. \ref{fig: robot_arm}). Therefore, the configuration space $\manifold=\mathbb{T}^2=\mathbb{S}^1 \times \mathbb{S}^1$ is a torus. We aim to let the joint angles follow the desired path $\set{P} \subseteq \manifold$ in the joint space as below:
	\begin{equation} \label{eq_pathtorus}
		\set{P} = \{(\theta_1, \theta_2) \in \mathbb{S}^1 \times \mathbb{S}^1: \phi(\theta_1, \theta_2) = 0 \},
	\end{equation}
	where $\phi(\theta_1, \theta_2) = \theta_1+\theta_2-\pi/2$. The desired path in the joint space described by these two angles corresponds to a circle trajectory of the center of the end-effector in the Cartesian space. In other words, if the joint angles $\theta_1$ and $\theta_2$ are controlled to follow $\set{P}$, then it turns out that the center of the end-effector will follow a circle centered at $(0, L_2)$ with radius $L_1$, where $L_1$ and $L_2$ are the link lengths of the robot arm. The configuration space $\manifold$ and the desired path $\set{P} \subseteq \manifold$ are illustrated in Fig. \ref{fig:torusrobotarm}. Since $\set{M}=\mathbb{T}^2$ is not homotopy equivalent to $\mathbb{S}^1$, global convergence from $\manifold$ to the desired path $\set{P} \subseteq \set{M}$ is not possible from Theorem \ref{thm_homotopy}. However, one way to achieve global convergence\footnote{More precisely, this global convergence is considered from the covering space $\mbr[2]$ rather than $\set{M}=\mathbb{T}^2$.} is to lift the problem to the covering space\cite[pp. 278-287]{lee2010topologicalmanifolds} $\mathbb{R}^2$ of $\mathbb{T}^2$ and the simple-closed desired path $\set{P}$ is thereby transformed to a straight line in $\mbr[2]$, which is a deformation retract of $\mbr[2]$. Then the vector-field-guided path-following problem can be solved in the space $\mbr[2]$. More specific, we regard the angles in \eqref{eq_pathtorus} as $(\theta_1, \theta_2) \in \mbr[2]$, and the function $\phi: \mbr[2] \to \mbr[]$, and derive the guiding vector field using \eqref{gvfe}. Here, the vector field is no longer defined on $\mathbb{T}^2$ but on $\mbr[2]$. By some computations, one can observe that there are no singular points of the vector field \eqref{gvfe} in $\mbr[2]$, and therefore, the global convergence from any point $(\theta_1, \theta_2) \in \mbr[2]$ in the covering space to the desired path $\set{P}$ is expected. As seen from Fig. \ref{fig:torusrobotarm}, three integral curves of the vector field \eqref{gvfe} all converge to the desired path $\set{P}$. For more detail about the idea of ``lifting'' the original configuration space to achieve global convergence to the desired path in the Euclidean space, see \cite{yao2020singularity}.

\begin{figure}
		\centering
		\includegraphics[width=0.8\linewidth]{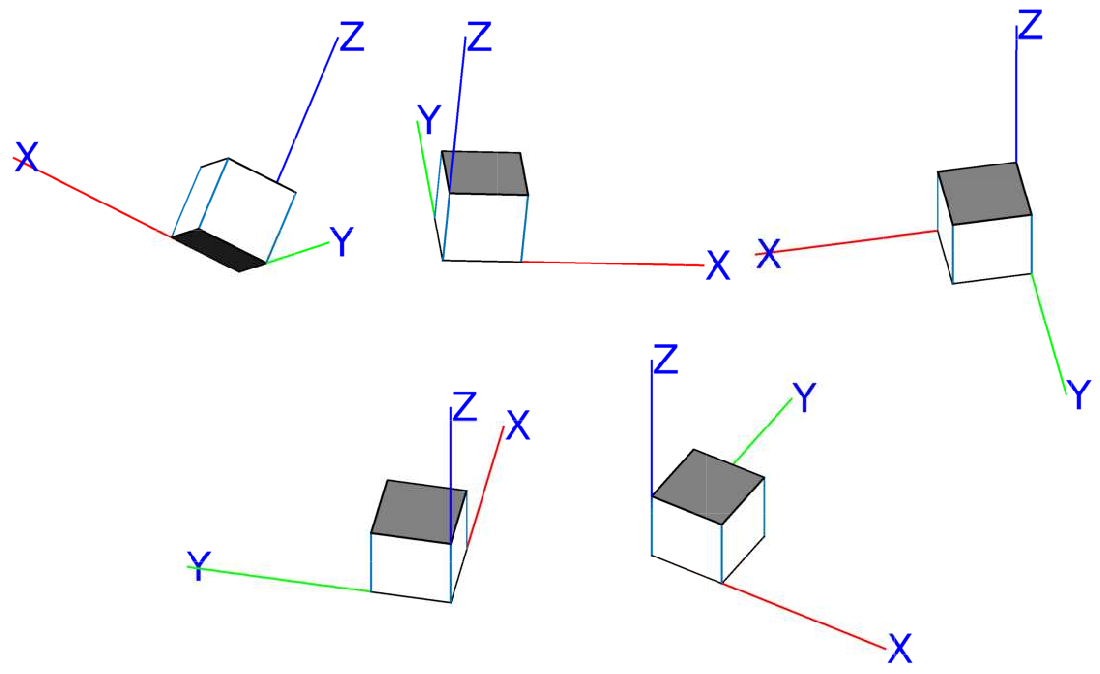}
		\caption{The motions of the coordinate frames, where cubes are used for better visualization. The coordinate frames at time instants 0.0, 1.4, 4.1, 6.9 and 9.6 seconds are shown from left to right, top to bottom. }
		\label{fig:so3frames}
	\end{figure}

	In the fifth example, the manifold $\manifold$ is the special orthogonal group $SO(3)$, which consists of a set of orthogonal matrices whose determinants are $1$. Since $SO(3)$ is an  embedded submanifold of $\mbr[3 \times 3]$, we have $SO(3) = \{ A \in \mbr[3 \times 3] : A^\top A = I, \det A = 1\}$. Therefore, we can choose six functions in \eqref{eq_manifold_euclidean} as follows: $f_1=A_1^\top A_2$,  $f_2=A_1^\top A_3$, $f_3=A_2^\top A_3$, $f_4=A_1^\top A_1-1$, $f_5=A_2^\top A_2-1$, $f_6=A_3^\top A_3-1$, and $a_i=0$ for $i=1,\dots,6$, in  \eqref{eq_manifold_euclidean}, where $A_j$ is the $j$-th column of the matrix $A$, for $j=1,2,3$. \footnote{These constraints do not rule out the possibility that $\det A=-1$, but once the initial configuration is in $SO(3)$, then the whole trajectory is always in $SO(3)$; i.e., the determinant of any matrix of the trajectory is always $1$. This is due to the fact that $\{A \in O(3): \det A = -1\}$ and $\{A \in O(3): \det A = 1\}=SO(3)$ are disjoint. } Note that the gradient of the function $f_i$, $i=1,\dots,6$, is a column vector consisting of the partial derivatives of $f_i$ with respect to each of the nine entries of the matrix $A \in \mbr[3 \times 3]$. We choose the desired path:
	\begin{equation} \label{eq_example5}
	\begin{split}
			\set{P} &= \{ A \in SO(3): \phi_1(A)=a_{13}=0, \phi_2(A)=a_{23}=0\} \\
			&= \underbrace{ \{ \Rot{z}{\theta} : \theta \in \mathbb{S}^1\} }_{\set{P}_1} \cup \underbrace{ \{ \Rot{x}{\pi} \Rot{z}{\theta} : \theta \in \mathbb{S}^1\} }_{\set{P}_2},
	\end{split}
	\end{equation}
	where $a_{ij}$ is the $ij$-th entry of the matrix $A$, and $\Rot{\{x,y,z\}}{\theta} \in SO(3)$ is the rotation matrix encoding the rotation of $\theta$ rads about the $x$, $y$, or $z$-axis. Since $\set{P}$ constitutes two disjoint components $\set{P}_1$ and $\set{P}_2$ in \eqref{eq_example5}, which component a trajectory converges to relies on the initial condition. We choose the initial condition to be $\xi_0 = \Rot{x}{\pi/4}\Rot{y}{-\pi/4} \in SO(3)$, and the corresponding trajectory converges to $\set{P}_1$. Intuitively, the trajectory in $SO(3)$ represents continuous pose transitions starting from $\Rot{x}{\pi/4}\Rot{y}{-\pi/4}$ to rotations about the $z$-axis of the ``identity pose'' $I=\Rot{x}{0}\Rot{y}{0}\Rot{z}{0}$ (see Fig. \ref{fig:so3frames}). 
\begin{remark}
	It might be elusive to design a desired path $\set{P}$ on $SO(3)$, which is the zero regular level set of the ``stacked'' function $\Phi \defeq (\phi_1, \phi_2): SO(3) \to \mbr[2]$. One convenient approach is to introduce two continuous surjective functions $\mu: SO(3) \to \mathbb{S}^2 \subseteq \mbr[3]$ defined by  $R \mapsto R v$ with $v=(0,0,1) \in \mathbb{S}^2 \subseteq \mbr[3]$, and $\pi_{xy}: \mathbb{S}^2 \subseteq \mbr[3] \to \mbr[2]$ being the projection of the first two coordinates onto the $xy$-plane. One can check that $\mu$ is a submersion on $SO(3)$ and $\pi_{xy}$ is a submersion on $\mathbb{S}^2 \setminus \{(x,y,z) \in \mathbb{S}^2 \subseteq \mbr[3] : z = 0 \}$. Let $\Phi = \pi_{xy} \circ \mu$; then every point of the open disk $\{(x,y) \in \mbr[2]: x^2+y^2<1\}$ is a regular point of $\Phi$. Therefore, the desired path $\set{P}$ above is the (regular) zero-level set $\Phi^{-1}\big( (0,0) \big)= (\pi_{xy} \circ \mu)^{-1} \big( (0,0) \big)=\mu^{-1}\big((0,0,1) \big) \cup \mu^{-1}\big((0,0,-1) \big) = \set{P}_1 \cup \set{P}_2$ in \eqref{eq_example5}.  It is of interest to investigate more complicated desired paths in $SO(3)$ (i.e., pose motions) by specifying different functions $\Phi$.
\end{remark}
	
	\section{Conclusions and future work}  \label{sec_conclude}
	We investigate some topological aspects of a guiding vector field defined on a general smooth Riemannian manifold for the path-following control problem, and analyze the stability and attractiveness of the desired path and the singular set. Specifically, we first generalize the widely-studied guiding vector field concept defined on Euclidean space to one defined on a general smooth manifold such that the new guiding vector field enables asymptotic following of a desired path defined on a manifold, such as a torus for robot arm joint space control applications. Then, motivated by many examples in the literature, we propose a question regarding whether singular points always exist in a guiding vector field when the desired path is homeomorphic to the unit circle. This question is directly related to the possibility of global convergence to the desired path. Since we consider a general smooth $n$-dimensional Riemannian manifold $\manifold$, the Poincar{\' e}-Bendixson theorem is not always applicable. 
	
	To answer this question, we first derive the dichotomy convergence property of the new guiding vector field, and conduct stability and attractiveness analysis of the desired path and the singular set. Then we revisit an existing topological result and provide some interpretations and implications for the vector field guided path-following problem. It turns out that the domain of attraction of the desired path (homeomorphic to the unit circle)  is homotopy equivalent to the unit circle. For the particular case when $\manifold=\mbr[n]$, we conclude that global convergence to the desired path is impossible, and singular points always exist when the dichotomy convergence property holds globally. Motivated by this impossibility result of global convergence to the desired path, we have shown the existence of non-path-converging trajectories. More specifically, we have proved that for any ball containing the desired path, there always exists at least one non-path-converging trajectory starting from the boundary of the ball. This result is related to the topological aspect of motion planning problems in obstacle-populated environments. Several numerical examples are provided to illustrate the theoretical results. 
	
	One of our results' implications is that one should retreat to other approaches if global convergence in $\mbr[n]$ is required. A possible solution is to change the topology of the desired path by ``tearing'' and ``stretching'' it along a virtual axis, and thus the new desired path is unbounded and homeomorphic to the real line in a possibly higher-dimensional space \cite{yao2020mobile2,yao2020singularity}.  The
	detailed proofs of Conjectures \ref{conjecture1} and \ref{conjecture2} are left for future work.
	
	\bibliographystyle{IEEEtran}
	\bibliography{ref}

\begin{thebibliography}{10}
\providecommand{\url}[1]{#1}
\csname url@samestyle\endcsname
\providecommand{\newblock}{\relax}
\providecommand{\bibinfo}[2]{#2}
\providecommand{\BIBentrySTDinterwordspacing}{\spaceskip=0pt\relax}
\providecommand{\BIBentryALTinterwordstretchfactor}{4}
\providecommand{\BIBentryALTinterwordspacing}{\spaceskip=\fontdimen2\font plus
\BIBentryALTinterwordstretchfactor\fontdimen3\font minus
  \fontdimen4\font\relax}
\providecommand{\BIBforeignlanguage}[2]{{%
\expandafter\ifx\csname l@#1\endcsname\relax
\typeout{** WARNING: IEEEtran.bst: No hyphenation pattern has been}%
\typeout{** loaded for the language `#1'. Using the pattern for}%
\typeout{** the default language instead.}%
\else
\language=\csname l@#1\endcsname
\fi
#2}}
\providecommand{\BIBdecl}{\relax}
\BIBdecl

\bibitem{aguiar2005path}
A.~P. Aguiar, J.~P. Hespanha, and P.~V. Kokotovic, ``Path-following for
  nonminimum phase systems removes performance limitations,'' \emph{IEEE
  Transactions on Automatic Control}, vol.~50, no.~2, pp. 234--239, 2005.

\bibitem{aguiar2008performance}
A.~P. Aguiar, J.~P. Hespanha, and P.~V. Kokotovi{\'c}, ``Performance
  limitations in reference tracking and path following for nonlinear systems,''
  \emph{Automatica}, vol.~44, no.~3, pp. 598--610, 2008.

\bibitem{Sujit2014}
P.~B. Sujit, S.~Saripalli, and J.~B. Sousa, ``Unmanned aerial vehicle path
  following: A survey and analysis of algorithms for fixed-wing unmanned aerial
  vehicless,'' \emph{IEEE Control Systems}, vol.~34, no.~1, pp. 42--59, Feb
  2014.

\bibitem{Goncalves2010}
V.~M. Goncalves, L.~C.~A. Pimenta, C.~A. Maia, B.~C.~O. Dutra, and G.~A.~S.
  Pereira, ``Vector fields for robot navigation along time-varying curves in
  $n$-dimensions,'' \emph{IEEE Transactions on Robotics}, vol.~26, no.~4, pp.
  647--659, Aug 2010.

\bibitem{lawrence2008lyapunov}
D.~A. Lawrence, E.~W. Frew, and W.~J. Pisano, ``Lyapunov vector fields for
  autonomous unmanned aircraft flight control,'' \emph{Journal of Guidance,
  Control, and Dynamics}, vol.~31, no.~5, pp. 1220--1229, 2008.

\bibitem{yao2020auto}
W.~Yao and M.~Cao, ``Path following control in {3D} using a vector field,''
  \emph{Automatica}, vol. 117, p. 108957, 2020.

\bibitem{yao2021thesis}
W.~Yao, ``Guiding vector fields for robot motion control,'' Ph.D. dissertation,
  University of Groningen, 2021.

\bibitem{phillips2004mechanics}
W.~F. Phillips, \emph{Mechanics of flight}.\hskip 1em plus 0.5em minus
  0.4em\relax John Wiley \& Sons, 2004.

\bibitem{kapitanyuk2017guiding}
Y.~A. Kapitanyuk, A.~V. Proskurnikov, and M.~Cao, ``A guiding vector-field
  algorithm for path-following control of nonholonomic mobile robots,''
  \emph{IEEE Transactions on Control Systems Technology}, vol.~26, no.~4, pp.
  1372--1385, July 2018.

\bibitem{rezende2018robust}
A.~M. Rezende, V.~M. Gon{\c{c}}alves, G.~V. Raffo, and L.~C. Pimenta, ``Robust
  fixed-wing {UAV} guidance with circulating artificial vector fields,'' in
  \emph{2018 IEEE/RSJ International Conference on Intelligent Robots and
  Systems (IROS)}.\hskip 1em plus 0.5em minus 0.4em\relax IEEE, 2018, pp.
  5892--5899.

\bibitem{lakomy2017}
K.~{\L}akomy and M.~M. Micha{\l}ek, ``The {VFO} path-following kinematic
  controller for robotic vehicles moving in a {3D} space,'' in \emph{Robot
  Motion and Control (RoMoCo), 2017 11th International Workshop on}.\hskip 1em
  plus 0.5em minus 0.4em\relax IEEE, 2017, pp. 263--268.

\bibitem{nelson2007vector}
D.~R. Nelson, D.~B. Barber, T.~W. McLain, and R.~W. Beard, ``Vector field path
  following for miniature air vehicles,'' \emph{IEEE Transactions on Robotics},
  vol.~23, no.~3, pp. 519--529, 2007.

\bibitem{Zhu2013}
S.~Zhu, D.~Wang, and C.~B. Low, ``Ground target tracking using {UAV} with input
  constraints,'' \emph{Journal of Intelligent {\&} Robotic Systems}, vol.~69,
  no.~1, pp. 417--429, Jan 2013.

\bibitem{kapitanyuk2017guiding2}
Y.~A. Kapitanyuk, H.~G. de~Marina, A.~V. Proskurnikov, and M.~Cao, ``Guiding
  vector field algorithm for a moving path following problem,''
  \emph{IFAC-PapersOnLine}, vol.~50, no.~1, pp. 6983--6988, 2017.

\bibitem{lee2010topologicalmanifolds}
J.~Lee, \emph{Introduction to topological manifolds}.\hskip 1em plus 0.5em
  minus 0.4em\relax Springer Science \& Business Media, 2010, vol. 202.

\bibitem{liang2016combined}
Y.~Liang and Y.~Jia, ``Combined vector field approach for {2D} and {3D}
  arbitrary twice differentiable curved path following with constrained
  {UAV}s,'' \emph{Journal of Intelligent \& Robotic Systems}, vol.~83, no.~1,
  pp. 133--160, 2016.

\bibitem{michalek2018vfo}
M.~M. Micha{\l}ek and T.~Gawron, ``{VFO} path following control with guarantees
  of positionally constrained transients for unicycle-like robots with
  constrained control input,'' \emph{Journal of Intelligent \& Robotic
  Systems}, vol.~89, no. 1-2, pp. 191--210, 2018.

\bibitem{fossen2003line}
T.~I. Fossen, M.~Breivik, and R.~Skjetne, ``Line-of-sight path following of
  underactuated marine craft,'' \emph{IFAC Proceedings Volumes}, vol.~36,
  no.~21, pp. 211--216, 2003.

\bibitem{yao2018cdc}
W.~Yao, Y.~A. Kapitanyuk, and M.~Cao, ``Robotic path following in {3D} using a
  guiding vector field,'' in \emph{IEEE Conference on Decision and Control},
  2018, pp. 4475--4480.

\bibitem{bullo2019geometric}
F.~Bullo and A.~D. Lewis, \emph{Geometric control of mechanical systems:
  modeling, analysis, and design for simple mechanical control systems}.\hskip
  1em plus 0.5em minus 0.4em\relax Springer, 2019, vol.~49.

\bibitem{lavalle2006planning}
S.~M. LaValle, \emph{Planning algorithms}.\hskip 1em plus 0.5em minus
  0.4em\relax Cambridge university press, 2006.

\bibitem{koditschek1990robot}
D.~E. Koditschek and E.~Rimon, ``Robot navigation functions on manifolds with
  boundary,'' \emph{Advances in applied mathematics}, vol.~11, p. 412, 1990.

\bibitem{lee2015introduction}
J.~Lee, \emph{Introduction to Smooth Manifolds: Second Edition}, ser. Graduate
  texts in mathematics.\hskip 1em plus 0.5em minus 0.4em\relax Springer, 2015.

\bibitem{nijmeijer1990nonlinear}
H.~Nijmeijer and A.~Van~der Schaft, \emph{Nonlinear dynamical control
  systems}.\hskip 1em plus 0.5em minus 0.4em\relax Springer, 1990, vol. 175.

\bibitem{lee2018introduction}
J.~M. Lee, \emph{Introduction to Riemannian manifolds}.\hskip 1em plus 0.5em
  minus 0.4em\relax Springer, 2018.

\bibitem{galbis2012vector}
A.~Galbis and M.~Maestre, \emph{Vector analysis versus vector calculus}.\hskip
  1em plus 0.5em minus 0.4em\relax Springer Science \& Business Media, 2012.

\bibitem{axler2015linear}
S.~Axler, \emph{Linear algebra done right}.\hskip 1em plus 0.5em minus
  0.4em\relax Springer, 2015.

\bibitem{caharija2016integral}
W.~Caharija, K.~Y. Pettersen, M.~Bibuli, P.~Calado, E.~Zereik, J.~Braga, J.~T.
  Gravdahl, A.~J. S{\o}rensen, M.~Milovanovi{\'c}, and G.~Bruzzone, ``Integral
  line-of-sight guidance and control of underactuated marine vehicles: Theory,
  simulations, and experiments,'' \emph{IEEE Transactions on Control Systems
  Technology}, vol.~24, no.~5, pp. 1623--1642, 2016.

\bibitem{yao2019integrated}
W.~Yao, B.~Lin, and M.~Cao, ``Integrated path following and collision avoidance
  using a composite vector field,'' in \emph{2019 IEEE 58th Conference on
  Decision and Control (CDC)}.\hskip 1em plus 0.5em minus 0.4em\relax IEEE,
  2019, pp. 250--255.

\bibitem{yao2021collision}
W.~Yao, B.~Lin, B.~D.~O. Anderson, and M.~Cao, ``Guiding vector fields for
  following occluded paths,'' \emph{IEEE Transactions on Automatic Control
  (TAC)}, 2021, conditionally accepted.

\bibitem{yao2021dichotomy}
------, ``Refining dichotomy convergence in vector-field guided path following
  control,'' in \emph{European Control Conference (ECC)}, 2021.

\bibitem{helmke2012optimization}
U.~Helmke and J.~B. Moore, \emph{Optimization and dynamical systems}.\hskip 1em
  plus 0.5em minus 0.4em\relax Springer Science \& Business Media, 2012.

\bibitem{fraleigh1995linear}
J.~B. Fraleigh and A.~B. Raymond, ``Linear algebra 3rd ed,'' \emph{Reading:
  Addison Wesley}, 1995.

\bibitem{khalil2002nonlinear}
H.~Khalil, \emph{Nonlinear Systems}, 3rd~ed.\hskip 1em plus 0.5em minus
  0.4em\relax Prentice Hall, 2002.

\bibitem{haddad2011nonlinear}
W.~M. Haddad and V.~Chellaboina, \emph{Nonlinear dynamical systems and control:
  a Lyapunov-based approach}.\hskip 1em plus 0.5em minus 0.4em\relax Princeton
  university press, 2011.

\bibitem{bhatia2002stability}
N.~P. Bhatia and G.~P. Szeg{\"o}, \emph{Stability theory of dynamical
  systems}.\hskip 1em plus 0.5em minus 0.4em\relax Springer Science \& Business
  Media, 2002.

\bibitem{sastry2013nonlinear}
S.~Sastry, \emph{Nonlinear systems: analysis, stability, and control}.\hskip
  1em plus 0.5em minus 0.4em\relax Springer Science \& Business Media, 2013,
  vol.~10.

\bibitem{moulay2010topological}
E.~Moulay and S.~P. Bhat, ``Topological properties of asymptotically stable
  sets,'' \emph{Nonlinear Analysis: Theory, Methods \& Applications}, vol.~73,
  no.~4, pp. 1093--1097, 2010.

\bibitem{wilson1969smoothing}
F.~W. Wilson, ``Smoothing derivatives of functions and applications,''
  \emph{Transactions of the American Mathematical Society}, vol. 139, pp.
  413--428, 1969.

\bibitem{wilson1967structure}
F.~Wilson~Jr, ``The structure of the level surfaces of a lyapunov function,''
  \emph{Journal of differential equations}, vol.~3, no.~3, pp. 323--329, 1967.

\bibitem{conley1978isolated}
C.~C. Conley, \emph{Isolated invariant sets and the Morse index}.\hskip 1em
  plus 0.5em minus 0.4em\relax American Mathematical Soc., 1978, no.~38.

\bibitem{yao2021doa}
W.~Yao, B.~Lin, B.~D.~O. Anderson, and M.~Cao, ``The domain of attraction of
  the desired path in vector-field guided path following,'' 2021, submitted.

\bibitem{yao2020singularity}
\BIBentryALTinterwordspacing
W.~Yao, H.~G. de~Marina, B.~Lin, and M.~Cao, ``Singularity-free guiding vector
  field for robot navigation,'' \emph{IEEE Transactions on Robotics}, 2021.
  [Online]. Available: \url{https://arxiv.org/abs/2012.01826}
\BIBentrySTDinterwordspacing

\bibitem{yao2020mobile2}
W.~Yao, H.~G. de~Marina, and M.~Cao, ``Vector field guided path following
  control: Singularity elimination and global convergence,'' in \emph{2020 IEEE
  59th Conference on Decision and Control (CDC)}, 2020.

\bibitem{guillemin2010differential}
V.~Guillemin and A.~Pollack, \emph{Differential topology}.\hskip 1em plus 0.5em
  minus 0.4em\relax American Mathematical Soc., 2010, vol. 370.

\bibitem{bing1983geometric}
R.~H. Bing, \emph{The geometric topology of 3-manifolds}.\hskip 1em plus 0.5em
  minus 0.4em\relax American Mathematical Soc., 1983, vol.~40.

\bibitem{brown1960proof}
M.~Brown, ``A proof of the generalized schoenflies theorem,'' \emph{Bulletin of
  the American Mathematical Society}, vol.~66, no.~2, pp. 74--76, 1960.

\bibitem{mazur1959embeddings}
B.~Mazur, ``On embeddings of spheres,'' \emph{Bulletin of the American
  Mathematical Society}, vol.~65, no.~2, pp. 59--65, 1959.

\bibitem{RimonKoditschek1992}
E.~{Rimon} and D.~E. {Koditschek}, ``Exact robot navigation using artificial
  potential functions,'' \emph{IEEE Transactions on Robotics and Automation},
  vol.~8, no.~5, pp. 501--518, 1992.

\bibitem{Braun2017OnE}
\BIBentryALTinterwordspacing
P.~Braun and C.~Kellett, ``On (the existence of) {C}ontrol {L}yapunov {B}arrier
  {F}unctions,'' 2017. [Online]. Available:
  \url{https://epub.uni-bayreuth.de/3522/1/CLBFs_submission_pbraun.pdf}
\BIBentrySTDinterwordspacing

\end{thebibliography}
	
	\begin{IEEEbiography}[{\includegraphics[width=1in,height=1.25in,clip,keepaspectratio]{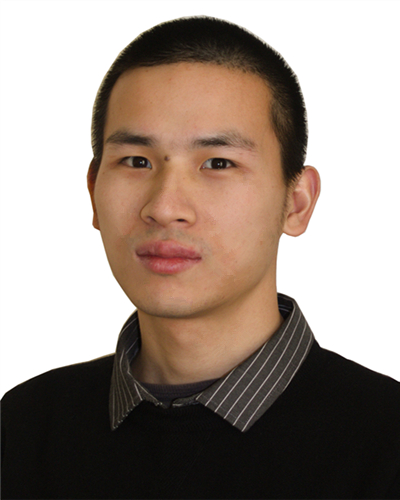}}]{Weijia Yao} obtained the Ph.D. degree with the distinction \textit{cum laude} in systems and control theory from the University of Groningen, Groningen, the Netherlands, in 2021.
		
	His research interests include nonlinear systems and control, robotics and multiagent systems.
	
	Dr. Yao was a finalist for the Best Conference Paper Award at ICRA in 2021, and the recipient of the Outstanding Master Degree Dissertation award of Hunan province, China, in 2020.
	\end{IEEEbiography}

	\begin{IEEEbiography}[{\includegraphics[width=1in,height=1.25in,clip,keepaspectratio]{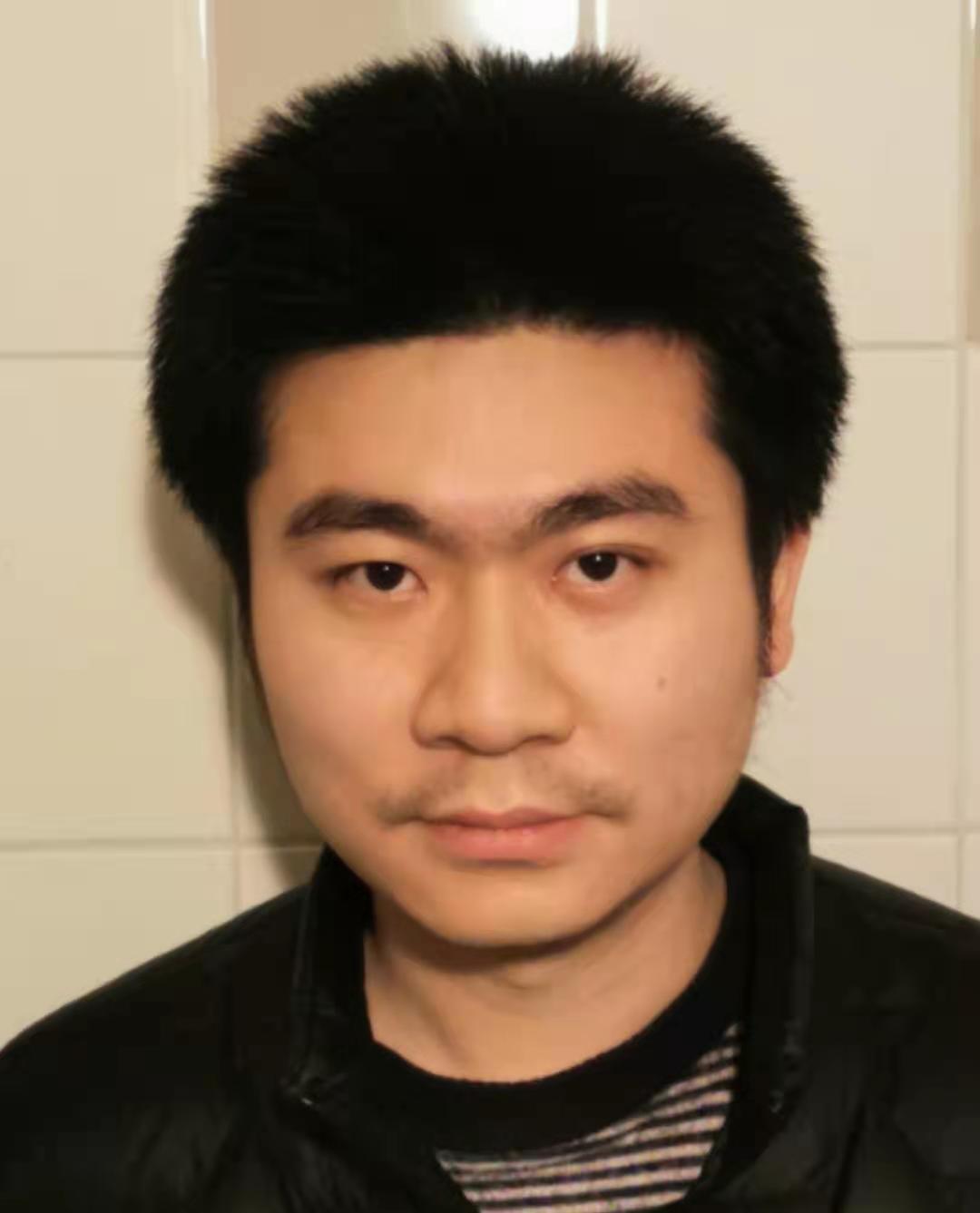}}]{Bohuan Lin} got
	 his BSc and MSc from Sun Yat-sen University. He is now working on a PhD project on integrable systems at the Bernoulli institute of University of Groningen. His research interests currently focus on topological/geometric aspects of dynamical systems.
	\end{IEEEbiography}

	\begin{IEEEbiography}[{\includegraphics[width=1in,height=1.25in,clip,keepaspectratio]{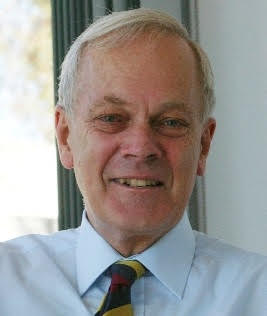}}]{Brian D. O. Anderson} (M’66-SM’74-F’75-LF’07) was born in Sydney, Australia, and educated at Sydney University in mathematics and electrical engineering, with PhD in electrical engineering from Stanford University in 1966.   He is an Emeritus Professor at the Australian National University (having retired as Distinguished Professor in 2016). His awards include the IEEE Control Systems Award of 1997, the 2001 IEEE James H Mulligan, Jr Education Medal, and the Bode Prize of the IEEE Control System Society in 1992, as well as several IEEE and other best paper prizes. He is a Fellow of the Australian Academy of Science, the Australian Academy of Technological Sciences and Engineering, the Royal Society, and a foreign member of the US National Academy of Engineering. He holds honorary doctorates from a number of universities, including Université Catholique de Louvain, Belgium, and ETH, Zürich. He is a past president of the International Federation of Automatic Control and the Australian Academy of Science. His current research interests are in distributed control, localization and social networks.
	\end{IEEEbiography}

	\begin{IEEEbiography}[{\includegraphics[width=1in,height=1.25in,clip,keepaspectratio]{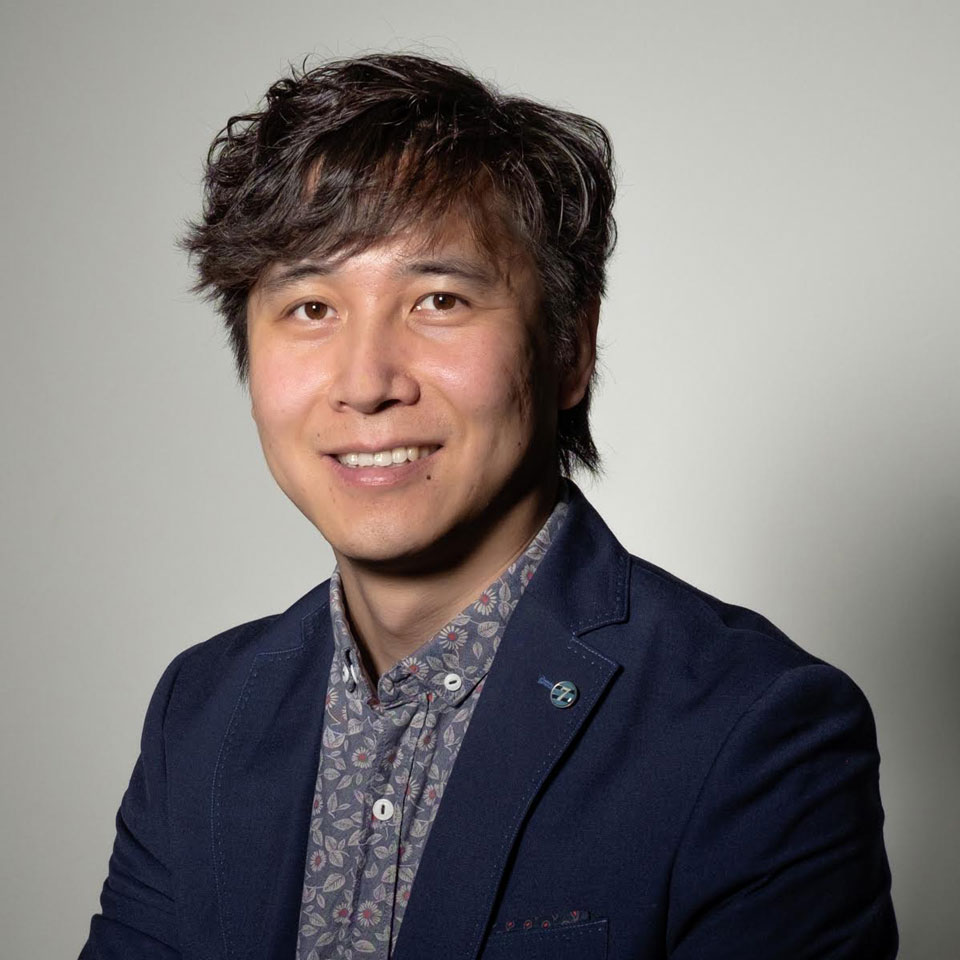}}]{Ming Cao}  has since 2016 been a professor of systems and control with the Engineering and Technology Institute (ENTEG) at the University of Groningen, the Netherlands, where he started as a tenure-track Assistant Professor in 2008. He received the Bachelor degree in 1999 and the Master degree in 2002 from Tsinghua University, Beijing, China, and the Ph.D. degree in 2007 from Yale University, New Haven, CT, USA, all in Electrical Engineering. From September 2007 to August 2008, he was a Postdoctoral Research Associate with the Department of Mechanical and Aerospace Engineering at Princeton University, Princeton, NJ, USA. He worked as a research intern during the summer of 2006 with the Mathematical Sciences Department at the IBM T. J. Watson Research Center, NY, USA. He is the 2017 and inaugural recipient of the Manfred Thoma medal from the International Federation of Automatic Control (IFAC) and the 2016 recipient of the European Control Award sponsored by the European Control Association (EUCA). He is a Senior Editor for Systems and Control Letters, and an Associate Editor for IEEE Transactions on Automatic Control, IEEE Transactions on Circuits and Systems and IEEE Circuits and Systems Magazine. He is a vice chair of the IFAC Technical Committee on Large-Scale Complex Systems. His research interests include autonomous agents and multi-agent systems, complex networks and decision-making processes.   
	\end{IEEEbiography}
\vfill
	
\end{document}